\documentclass[graybox]{svmult}

\usepackage{type1cm}        
\usepackage{makeidx}         
\usepackage{graphicx}        
\usepackage{multicol}        
\usepackage[bottom]{footmisc}

\usepackage{newtxtext}       %
\usepackage{newtxmath}       

\makeindex             

\usepackage{url}
\usepackage{xspace}
\usepackage{wrapfig}
\usepackage{amsmath}
\usepackage{mathrsfs}
\usepackage{pifont}
\usepackage{subcaption}
\usepackage{tikz}
\usetikzlibrary{calc}
\usetikzlibrary{positioning}
\usetikzlibrary{automata}
\usetikzlibrary{shapes.multipart}
\usetikzlibrary{shadows}
\usetikzlibrary{shapes}
\usetikzlibrary{decorations.pathreplacing}
\usetikzlibrary{backgrounds}
\usepackage{pgfplots}

\usepackage{listings}
\newcommand{\showcmt}[1]{#1}
\newcommand{\mtcmt}[1]{\showcmt{\textbf{\color{magenta}{}}}}
\newcommand{\dkcmt}[1]{\showcmt{\textbf{\color{orange}{}}}}
\newcommand{\pscmt}[1]{\showcmt{\textbf{\color{blue}{}}}}

\makeatletter
\let\UrlSpecialsOld\UrlSpecials
\def\UrlSpecials{\UrlSpecialsOld\do\/{\Url@slash}\do\_{\Url@underscore}}%
\def\Url@slash{\@ifnextchar/{\kern-.11em\mathchar47\kern-.2em}%
    {\kern-.0em\mathchar47\kern-.08em\penalty\UrlBigBreakPenalty}}
\def\Url@underscore{\nfss@text{\leavevmode \kern.06em\vbox{\hrule\@width.3em}}}
\makeatother

\newcounter{thecmd}
\newenvironment{cmd}{\refstepcounter{thecmd}\medskip\noindent\small\hspace*{1em}\tt}{\normalsize\rm\hfill C\arabic{thecmd}\medskip\noindent}
\newcommand{\refcmd}[1]{C\ref{#1}}

\begin{document}

\title*{CBMC}
\subtitle{The C Bounded Model Checker}

\author{Daniel Kroening, Peter Schrammel and Michael Tautschnig}
\institute{
Daniel Kroening \at University of Oxford, United Kingdom and Diffblue Ltd, Oxford, United Kingdom \email{kroening@cs.ox.ac.uk}
\and Peter Schrammel \at University of Sussex, Brighton, United Kingdom and Diffblue Ltd, Oxford, United Kingdom \email{p.schrammel@sussex.ac.uk}
\and Michael Tautschnig \at Queen Mary University of London, United Kingdom \email{michael.tautschnig@qmul.ac.uk}}

\maketitle


\section{Introduction} \label{sec:introduction}


The C Bounded Model Checker (CBMC)~\cite{DBLP:conf/tacas/ClarkeKL04}
demonstrates the violation of assertions in C programs, or proves
safety of the assertions under a given bound. CBMC implements a
bit-precise translation of an input C program, annotated with
assertions and with loops unrolled to a given depth, into a
formula. If the formula is satisfiable, then an execution leading to a
violated assertion exists.

CBMC is one of the most successful software verification tools.  Its
main advantages are its precision, robustness and simplicity.
CBMC is shipped as part of several Linux distributions.
It has been used by
thousands of software developers to verify real-world software, such
as the Linux kernel, and powers commercial software analysis and
test generation tools.
Table~\ref{tab:features} gives an overview of CBMC's features.

CBMC is also a versatile tool that can be applied to solve many
practical program analysis problems such as
bug finding, property checking, test input generation,
detection of security vulnerabilities, equivalence checking
and program synthesis.

This chapter will give an introduction into CBMC, including
practical examples and pointers to further reading.
Moreover, we give insights about the development of CBMC
itself, showing how its performance evolved over the last
decade.

Section~\ref{sec:approach} gives an overview
of the verification approach implemented by CBMC.
%
Section~\ref{sec:usecases} gives a tutorial on how to use
CBMC for various verification problems.
A strength of CBMC is its proven applicability to real-world
C programs; Section~\ref{sec:cprover} explains the features
that enable this.
Section~\ref{sec:architecture} describes the components of
CBMC's architecture.
Section~\ref{sec:history} gives an overview of how the performance
of CBMC evolved over the last 10 years, before wrapping up
in Section~\ref{sec:future}.

\begin{table}
  \begin{tabular}{|p{0.21\textwidth}|p{0.77\textwidth}|}
    \hline
    Languages  & C, GOTO \\
    \hline
    Properties & \textsf{assert}, memory safety, arithmetic overflow,
                 division-by-zero, memory leaks \\
    \hline
    Environments  & Linux, Mac OS, Windows, BSD \\
    \hline
    Technologies used  & symbolic execution, bounded model checking,
                         SAT and SMT solving \\
    \hline
    Other features & compilation and linking of entire projects into GOTO via \textsf{goto-cc}\\
    \hline
    Strengths  & memory safety, floating point arithmetic \\
    \hline
    Weaknesses & limited support for unbounded verification \\
    \hline
  \end{tabular}
  \caption{CBMC Features}
  \label{tab:features}
\end{table}


\noindent \emph{Software Project.}
The CPROVER framework (including CBMC) is implemented in C++
and has about 250\,KLOC.
CPROVER  is  maintained  by  Daniel  Kroening  with more than
60 contributors. It is made publicly available under a BSD-style license.
The source code and binaries for popular platforms are available at
\url{https://www.cprover.org/cbmc} and
\url{https://github.com/diffblue/cbmc}.
There is a detailed installation guide at
\url{https://www.cprover.org/cprover-manual/installation/}.

\section{Verification Approach} \label{sec:approach}


For a given Kripke structure, bounded model
checking~\cite{BCCZ99,handbook09} is a
semi-decision procedure that translates bounded unfoldings of a
transition relation and LTL formulae to propositional satisfiability
problems.  Soundness is achieved by incrementing the bound until a
witness is found, but completeness can only be achieved when the
number of steps to reach all states is finite.

CBMC implements bounded model checking for software, specifically for
C programs.
In this setting, the transition relation is specified by the C program
and the semantics laid out in the C language
standards~\cite{ANSI:1999:AII}, with the initial state determined by
the program's entry point.
As specification, the program is annotated with assertions rather than
using LTL formulae.
A bounded unfolding of the transition relation amounts to a bounded
number of execution steps of the program.
As a more practical notion of bounded unfolding, however, the bounded
unfolding is typically instantiated via bounded unrolling of loops and
recursive procedure calls.

Via such a bounded unfolding CBMC reduces questions about program
paths to constraints that can be solved by off-the-shelf Boolean
Satisfiability (SAT) or Satisfiability Modulo Theories
(SMT)~\cite{BarrettSST09} solvers.
With the SAT back end, and given a program annotated with assertions,
CBMC produces a CNF formula the solutions of which describe program
paths leading to assertion violations.
A model of this formula then amounts to a counterexample.

Before looking at the architecture in detail, let us consider an
example.
The program in Listing~\ref{prog:abs} shall be an attempt to compute
the absolute value of an integer-typed input.
\begin{lstlisting}[float=t, caption={Source code (\texttt{abs.c})}, label={prog:abs}]
int abs(int x) {
  int y = x;

  if(x < 0) {
    y = -x;
  }

  return y;
}
\end{lstlisting}
\begin{lstlisting}[float=t, language={}, caption={CBMC output of command~\refcmd{cmd:abs} for \texttt{abs.c} (Listing~\ref{prog:abs})}, label={output:abs}]
CBMC version 5.12 (cbmc-5.12-d8598f8-557-g1edf4d91f) 64-bit x86_64 macos *@\label{abs-out:l1}@*
Parsing abs.c *@\label{abs-out:l2}@*
Converting *@\label{abs-out:l3}@*
Type-checking abs *@\label{abs-out:l4}@*
Generating GOTO Program *@\label{abs-out:l5}@*
Adding CPROVER library (x86_64) *@\label{abs-out:l6}@*
Removal of function pointers and virtual functions *@\label{abs-out:l7}@*
Generic Property Instrumentation *@\label{abs-out:l8}@*
Running with 8 object bits, 56 offset bits (default) *@\label{abs-out:l9}@*
Starting Bounded Model Checking *@\label{abs-out:l10}@*
size of program expression: 45 steps *@\label{abs-out:l11}@*
simple slicing removed 0 assignments *@\label{abs-out:l12}@*
Generated 0 VCC(s), 0 remaining after simplification *@\label{abs-out:l13}@*
VERIFICATION SUCCESSFUL *@\label{abs-out:l14}@*
\end{lstlisting}
To run CBMC on this program, we need to specify the name of the source
file (\texttt{abs.c}) and the entry point \lstinline+abs+:

\begin{cmd}\label{cmd:abs}
cbmc -{}-function abs abs.c
\end{cmd}

\noindent with the output shown in Listing~\ref{output:abs}.
This output provides the following information:
\begin{itemize}
  \item Line~\ref{abs-out:l1} reports the version of CBMC being run (including the exact
    Git revision the CBMC executable was built from) and the platform
    it is running on.
  \item Lines~\ref{abs-out:l2}--\ref{abs-out:l5} report the source
    code being processed by the C front end of CBMC.
  \item Lines~\ref{abs-out:l6}--\ref{abs-out:l8} are status updates of
    the instrumentation steps.
  \item Line~\ref{abs-out:l9} confirms the (configurable) pointer
    encoding being used.
  \item Lines~\ref{abs-out:l10}--\ref{abs-out:l13} are status and
    statistics of symbolic execution.
    Notably, the input here caused zero verification conditions (VCCs)
    to be generated.
  \item Line~\ref{abs-out:l14} is CBMC's conclusive answer that no
    specification was violated.
    With the information in the preceding line (zero VCCs), however,
    this is vacuous: there were no assertions for CBMC to check.
    Indeed, the source code did not contain any \lstinline+assert+
    statements.
\end{itemize}
We could now either insert \lstinline+assert+ statements, or make use
of CBMC's built-in specifications.
In this case, arithmetic overflow is of particular interest.
Let us invoke CBMC again, with signed-integer overflow assertions
enabled:

\begin{cmd}\label{cmd:abs2}
  cbmc -{}-function abs -{}-signed-overflow-check abs.c
\end{cmd}

\noindent
We now obtain the output shown in Listing~\ref{output:abs2} (with the initial lines skipped):
\begin{lstlisting}[float=t, language={}, caption={CBMC output of command~\refcmd{cmd:abs2} for \texttt{abs.c} (Listing~\ref{prog:abs})}, label={output:abs2},firstnumber=10]
Starting Bounded Model Checking *@\label{abs-out-2:l10}@*
size of program expression: 46 steps *@\label{abs-out-2:l11}@*
simple slicing removed 7 assignments *@\label{abs-out-2:l12}@*
Generated 1 VCC(s), 1 remaining after simplification *@\label{abs-out-2:l13}@*
Passing problem to propositional reduction *@\label{abs-out-2:l14}@*
converting SSA *@\label{abs-out-2:l15}@*
Running propositional reduction *@\label{abs-out-2:l16}@*
Post-processing *@\label{abs-out-2:l17}@*
Solving with MiniSAT 2.2.1 with simplifier *@\label{abs-out-2:l18}@*
221 variables, 203 clauses *@\label{abs-out-2:l19}@*
SAT checker: instance is SATISFIABLE *@\label{abs-out-2:l20}@*
Runtime decision procedure: 0.00414769s *@\label{abs-out-2:l21}@*

** Results: *@\label{abs-out-2:l23}@*
abs.c function abs *@\label{abs-out-2:l23a}@*
[abs.overflow.1] line 5 arithmetic overflow on signed unary minus in -x: FAILURE *@\label{abs-out-2:l24}@*

** 1 of 1 failed (2 iterations) *@\label{abs-out-2:l27}@*
VERIFICATION FAILED *@\label{abs-out-2:l28}@*
\end{lstlisting}
Not only is the final verdict different (``VERIFICATION FAILED''), we
also see further differences:
\begin{itemize}
  \item Line~\ref{abs-out-2:l11} now reports one additional step, and
    line~\ref{abs-out-2:l13} confirms that a verification condition is
    now generated, i.e., we have a non-empty specification.
  \item An actual formula is being generated
    (lines~\ref{abs-out-2:l14}--\ref{abs-out-2:l17}).
    This formula is passed to and processed by a SAT solver
    (lines~\ref{abs-out-2:l18}--\ref{abs-out-2:l20}), which determines
    it to be satisfiable.
    As reported in line~\ref{abs-out-2:l21}, the SAT solver spent
    approximately 4\,ms to compute a model.
  \item The satisfiable formula amounts to a violated assertion, which
    is reported in lines~\ref{abs-out-2:l23}--\ref{abs-out-2:l27}.
    Specifically, an arithmetic overflow was detected
    (line~\ref{abs-out-2:l24}).
  \item The overall verdict is summarised in the last line of output.
\end{itemize}
The Boolean verification result and the one-line summary of the
violated specification typically are not sufficient for a software
engineer to debug the problem reported by CBMC.
As a model checker, CBMC internally computes a full counterexample.
In case of programs, a counterexample amounts to an execution trace.
CBMC will print the steps leading to a failing assertion with the
\texttt{-{}-trace} command-line option:

\begin{cmd}\label{cmd:abs3}
  cbmc -{}-function abs -{}-signed-overflow-check -{}-trace abs.c
\end{cmd}

\noindent
We now obtain the output shown in Listing~\ref{output:abs3} (again, initial lines are skipped).

\begin{lstlisting}[float=t, language={}, caption={CBMC output of command~\refcmd{cmd:abs3} for \texttt{abs.c} (Listing~\ref{prog:abs})}, label={output:abs3},firstnumber=23]
** Results: *@\label{abs-out-3:l23}@*
abs.c function abs *@\label{abs-out-3:l23a}@*
[abs.overflow.1] line 8 arithmetic overflow on signed unary minus in -x: FAILURE *@\label{abs-out-3:l24}@*

Trace for abs.overflow.1: *@\label{abs-out-3:l26}@*

State 17 file abs.c line 1 thread 0 *@\label{abs-out-3:l28}@*
---------------------------------------------------- *@\label{abs-out-3:l29}@*
  INPUT x: -2147483648 (10000000 00000000 00000000 00000000) *@\label{abs-out-3:l30}@*

State 20 file abs.c line 1 thread 0 *@\label{abs-out-3:l32}@*
---------------------------------------------------- *@\label{abs-out-3:l33}@*
  x=-2147483648 (10000000 00000000 00000000 00000000) *@\label{abs-out-3:l34}@*

State 21 file abs.c line 2 function abs thread 0 *@\label{abs-out-3:l36}@*
---------------------------------------------------- *@\label{abs-out-3:l37}@*
  y=0 (00000000 00000000 00000000 00000000) *@\label{abs-out-3:l38}@*

State 22 file abs.c line 2 function abs thread 0 *@\label{abs-out-3:l40}@*
---------------------------------------------------- *@\label{abs-out-3:l41}@*
  y=-2147483648 (10000000 00000000 00000000 00000000) *@\label{abs-out-3:l42}@*

Violated property: *@\label{abs-out-3:l44}@*
  file abs.c line 5 function abs *@\label{abs-out-3:l45}@*
  arithmetic overflow on signed unary minus in -x *@\label{abs-out-3:l46}@*
  !(x == -2147483648) *@\label{abs-out-3:l47}@*

** 1 of 1 failed (2 iterations) *@\label{abs-out-3:l51}@*
VERIFICATION FAILED *@\label{abs-out-3:l52}@*
\end{lstlisting}

CBMC now prints the input value (line~\ref{abs-out-3:l30}) that will
trigger an arithmetic overflow.
The value is printed both in decimal notation and in binary notation,
grouped as 8-bit bytes.
In this case the binary value is particularly insightful: this is the
maximum negative number representable when using two's complement over
32 bits.
The arithmetic overflow then occurs in line~5 of \texttt{abs.c} as
reported in lines~\ref{abs-out-3:l44}--\ref{abs-out-3:l47} of CBMC's
counterexample output.

CBMC implements the above steps following the pipeline outlined in
Figure~\ref{fig:arch} of Section~\ref{sec:architecture}.
In that section, we will break down each of CBMC's
components in detail to understand how CBMC arrives at its results.

%
%
%
%
%
%


\section{Using CBMC} \label{sec:usecases}




CBMC uses assertions to specify program properties. Assertions are
specifications over the state of the program when the program reaches a
particular program location. Assertions are often written by the
programmer using the \lstinline{assert} macro.

In addition to the assertions written by the programmer, assertions
for specific properties can also be generated automatically by CBMC,
often relieving the programmer from expressing properties that should
hold in any well-behaved program.
This assertion generator performs a conservative
static analysis to determine program locations that potentially
contain a bug. Due to the imprecision of the static analysis, it is
important to emphasise that these generated assertions are only
potential bugs, and that the model checker first needs to confirm that
they are indeed genuine bugs.

The assertion generator supports the subsequent verification
of the following properties:
\begin{itemize}
\item \emph{Buffer overflows}. For each array access, check whether
  the upper and lower bounds are violated.

\item \emph{Pointer safety}. Search for \lstinline{NULL}-pointer dereferences or
  dereferences of other invalid pointers.

\item \emph{Memory leaks}. Check whether the program constructs
  dynamically allocated data structures that are subsequently
  inaccessible.

\item \emph{Division by zero}. Check whether there is a division by
  zero in the program.

\item \emph{Not-a-Number}. Check whether floating-point computation
  may result in NaNs.

\item \emph{Arithmetic overflow}. Check whether a numerical overflow
  occurs during an arithmetic operation or type conversion.

\item \emph{Undefined shifts}. Check for shifts with excessive distance.
\end{itemize}


\noindent
All the properties described above are reachability properties. They
are always of the form
\textit{"Is there a path through the program such that some property is
violated?"}
The counterexamples to such properties are always program paths.
Stepping through these counterexamples is similar to debugging programs.

\subsection{Handling Loops}


As CBMC performs Bounded Model Checking, all loops have to have a
finite upper run-time bound in order to guarantee that all bugs are
found. CBMC can optionally check that sufficient unwinding is
performed.

As an example, consider the program \texttt{binsearch.c}
in Listing~\ref{prog:binsearch}.
\begin{lstlisting}[float=t, caption={Source code (\texttt{binsearch.c})}, label={prog:binsearch}]
int binsearch(int x)
{
  int a[16];
  signed low = 0, high = 16;

  while(low < high)
  {
    signed middle = low + ((high - low) >> 1); *@\label{binsearch:l8}@*

    if(a[middle] < x)
      high = middle;
    else if(a[middle] > x)
      low = middle + 1;
    else // a[middle]==x
      return middle;
  }

  return -1;
}
\end{lstlisting}
If you run CBMC on this function, you will notice that the unwinding
does not stop on its own. The built-in simplifier is not able to
determine a runtime bound for this loop. The unwinding bound has to
be given as a command line argument:

\begin{cmd}\label{cmd:binsearch}
\begin{tabular}{l@{ }l}
cbmc & binsearch.c -{}-function binsearch -{}-unwind 6 -{}-bounds-check\\
     & -{}-unwinding-assertions
\end{tabular}
\end{cmd}

\noindent The resulting output is shown in Listing~\ref{output:binsearch}.
CBMC verifies that the array accesses are within the bounds;
note that this actually depends on the result of the right shift in
Line~\ref{binsearch:l8} of the program. In
addition, as CBMC is given the option
\texttt{-{}-unwinding-assertions}, it also checks that sufficient
unwinding is done, i.e., it proves a runtime bound.

\begin{lstlisting}[float=t, language={}, caption={CBMC output of command~\refcmd{cmd:binsearch} for \texttt{binsearch.c} (Listing~\ref{prog:binsearch})}, label={output:binsearch}]
** Results:
binsearch.c function binsearch
[binsearch.unwind.0] line 6 unwinding assertion loop 0: SUCCESS
[binsearch.array_bounds.1] line 10
  array `a' lower bound in a[(signed long int)middle]: SUCCESS
[binsearch.array_bounds.2] line 10
  array `a' upper bound in a[(signed long int)middle]: SUCCESS
[binsearch.array_bounds.3] line 12 
  array `a' lower bound in a[(signed long int)middle]: SUCCESS
[binsearch.array_bounds.4] line 12 
  array `a' upper bound in a[(signed long int)middle]: SUCCESS
...
VERIFICATION SUCCESSFUL
\end{lstlisting}

For any lower
unwinding bound, there are traces that demonstrate that more loop
iterations are possible. Thus, CBMC will report that the unwinding assertion has
failed. As usual, a counterexample trace that documents this can be
obtained with the option \texttt{-{}-trace}.

CBMC can also be used for programs with unbounded loops. In this case,
CBMC is used for bug hunting only; CBMC does not attempt to find all
bugs. The program \texttt{lock.c} in Listing~\ref{prog:lock} is an example of a
program with a user-specified property.
\begin{figure}[!t]
\begin{multicols}{2}
\begin{lstlisting}
_Bool nondet_bool();
unsigned int nondet_unsigned_int();
_Bool LOCK = 0;

_Bool lock()
{
  if(nondet_bool())
  {
    assert(!LOCK);
    LOCK = 1;
    return 1;
  }

  return 0;
}

void unlock()
{
  assert(LOCK);
  LOCK = 0;
}
int main()
{
  unsigned got_lock = 0;
  unsigned times = nondet_unsigned_int();

  while(times > 0)
  {
    if(lock())
    {
      got_lock++;
      /* critical section */
    }

    if(got_lock != 0)
      unlock();

    got_lock--;
    times--;
  }
}
\end{lstlisting}
\end{multicols}
  \captionof{lstlisting}{Source code (\texttt{lock.c})}
  \label{prog:lock}
\end{figure}
The while loop in the main function has no (useful) runtime
bound. Thus, a bound has to be set on the amount of unwinding that
CBMC performs. There are two ways to do so:
\begin{enumerate}
\item The \texttt{-{}-unwind} command-line parameter can to be used to
  limit the number of times loops are unwound.
\item The \texttt{-{}-depth} command-line parameter can be used to limit
  the number of program steps to be processed.
\end{enumerate}

\noindent
For the example of Listing~\ref{prog:lock}, with a loop unwinding bound of one, no bug is
found. But for a bound of two, CBMC detects a trace that violates an
assertion. Without unwinding assertions, or when using the \texttt{-{}-depth}
command-line switch, CBMC does not necessarily prove the program correct, but it
can be helpful to find program bugs.
More information on limiting unwinding of loops can be found
at \url{https://www.cprover.org/cprover-manual/cbmc/unwinding/}.


\subsection{Using Built-in Checks}


\begin{lstlisting}[float=t, caption={Source code (\texttt{login.c})}, label={prog:security}]
#include <stdio.h>

int main (int argc, char** argv)
{
  char password[8] = {'s','e','c','r','e','t','!','\0'};
  char buffer[16] = {'\0', };
  int tmp;
  int index = 0;

  printf("Enter your name: ");
  while ((tmp = getchar()) != '\n')
  {
    buffer[index] = tmp;
    ++index;
  }

  printf("%s\n",buffer);

  return 0;
}
\end{lstlisting}

The issue of buffer overflows has obtained wide public attention. A
buffer is a contiguously allocated chunk of memory, represented by an
array or a pointer in C. Programs written in C do not provide
automatic bounds checking on the buffer, which means a program can --
accidentally or deliberately -- write beyond a buffer. The example
program in Listing~\ref{prog:security} is a syntactically valid C
program, compiling and executing (seemingly) without any errors.
If compiled on a system with the stack growing downwards, such as x86,
the following can be observed:
\begin{lstlisting}[language={}]
> gcc login.c -o login
> ./login
Enter your name: Daniel
Daniel
> ./login
Enter your name: Sim Sala Bim ...
Sim Sala Bim ...secret!
\end{lstlisting}

What has happened? The end of a character string in C is determined by
a \texttt{'\textbackslash0'} character. When we enter more than 15
characters then \lstinline{buffer} will not have any
\texttt{'\textbackslash0'} character at the end and \lstinline{printf}
will continue printing characters beyond the memory allocated for
\lstinline{buffer} until it encounters a \texttt{'\textbackslash0'}
character or crashes due to a memory access violation (segmentation
fault). Depending on the memory layout this might lead to disclosure
of confidential data. In our case above, the password is printed to
the terminal.

Could we have found this problem with the help of CBMC?
Yes, CBMC is able to check whether memory accesses beyond the bounds
of an allocated object are possible.
When we run

\begin{cmd}\label{cmd:security}
cbmc login.c -{}-unwind 20 -{}-bounds-check
\end{cmd}

\noindent CBMC reports
\begin{lstlisting}[language={}]
** Results:
login.c function main
[main.array_bounds.2] line 13
  array `buffer' upper bound in buffer[(signed long int)index]:
  FAILURE
[main.array_bounds.1] line 13
  array `buffer' lower bound in buffer[(signed long int)index]:
  SUCCESS
\end{lstlisting}
This means that the program is indeed faulty. Inspecting the trace
(\texttt{-{}-trace}) as shown in Listing~\ref{output:security} confirms that the problem occurs when
we assign to \lstinline{buffer} after \lstinline{index} has been incremented to 16.

\begin{lstlisting}[float=h, language={}, caption={CBMC output for command~\refcmd{cmd:security} with \texttt{-{}-trace} for \texttt{login.c} (Listing~\ref{prog:security})}, label={output:security}]
...
State 251 file login.c function main line 13 thread 0
----------------------------------------------------
  buffer[15l]=-1 (11111111)

State 252 file login.c function main line 14 thread 0
----------------------------------------------------
  index=16 (00000000 00000000 00000000 00010000)
...
State 259 file <builtin-library-getchar> function getchar line 15 thread 0
----------------------------------------------------
  INPUT getchar: -1 (11111111 11111111 11111111 11111111)
...
Violated property:
  file login.c function main line 13 thread 0
  array `buffer' upper bound in buffer[(signed long int)index]
  !((signed long int)index >= 16l)
\end{lstlisting}

If we enter further characters we would write beyond \lstinline{buffer}
and thus overwrite data on the stack.  In particular, such bugs can be
exploited to overwrite the return address of a function, thus enabling
the execution of arbitrary code.

CBMC is capable of detecting such bugs by checking these lower and
upper bounds, even for arrays with dynamic size:
The two options \texttt{-{}-bounds-check} and \texttt{-{}-pointer-check}
instruct CBMC to look for errors related to pointers and array
bounds.
When invoked with \texttt{-{}-show-properties}, CBMC will print the
list of properties it checks:

\begin{cmd}\label{cmd:runtime1}
cbmc login.c -{}-show-properties -{}-bounds-check -{}-pointer-check
\end{cmd}

\noindent
Note that it
lists, among others, a property labelled with ``array `buffer' upper bound'' together with the location of the faulty
array access:

\begin{lstlisting}[language={}]
Property main.array_bounds.1:
  file login.c line 13 function main
  array `buffer' lower bound in buffer[(signed long int)index]
  (signed long int)index >= 0l

Property main.array_bounds.2:
  file login.c line 13 function main
  array `buffer' upper bound in buffer[(signed long int)index]
  !((signed long int)index >= 16l)
\end{lstlisting}

As you can see, CBMC largely determines the property it
needs to check itself. This is realised by means of a preliminary
static analysis, which relies on computing a fixed point on various
abstract domains.
These automatically generated properties need not
necessarily correspond to bugs -- these are just potential flaws for
abstract interpretation might be imprecise. Whether these properties
hold or correspond to actual bugs needs to be determined by further
analysis.

CBMC performs this analysis using symbolic simulation, which
is facilitated by a translation of the program into a formula. The
formula is then combined with the property. Let's look at the formula
that is generated by CBMC's symbolic simulation:

\begin{cmd}\label{cmd:runtime2}
cbmc login.c -{}-unwind 20 -{}-show-vcc -{}-bounds-check -{}-pointer-check
\end{cmd}

\noindent
With this option, CBMC performs the symbolic simulation and prints the
verification conditions as a conjunction of equations.
A verification condition needs
to be proven to be valid by a decision procedure in order to assert
that the corresponding property holds. Let's run the decision
procedure:

\begin{cmd}\label{cmd:runtime3}
cbmc login.c -{}-unwind 20 -{}-bounds-check -{}-pointer-check
\end{cmd}

CBMC transforms the equation (that can be printed using
\texttt{-{}-show-vcc}) into CNF and passes
this formula to a SAT solver (cf.~\cite{DBLP:series/txtcs/KroeningS16}
for background on such transformations).
 It then determines which of the properties that
 it has generated for the program hold and which do not. Using the SAT
 solver, CBMC detects that the property for the object bounds of
 \lstinline{buffer}
 does not hold, and will display:

\begin{lstlisting}[language={}]
[main.array_bounds.2] line 13
  array `buffer' upper bound in buffer[(signed long int)index]:
  FAILURE
\end{lstlisting}

To aid the understanding of the problem, CBMC can generate a
counterexample trace for failed properties. To obtain this trace of
Listing~\ref{output:security}, run:

\begin{cmd}\label{cmd:runtime4}
cbmc login.c -{}-unwind 20 -{}-bounds-check -{}-pointer-check -{}-trace
\end{cmd}

\noindent
CBMC then prints a counterexample trace, that is, a program trace that
begins with main and ends in a state which violates the property. In
our example, the program trace ends in the faulty array access. It
also gives the values the input variables must have for the bug to
occur. In this example, the results of (repeated calls to)
\lstinline{getchar} must be ones to trigger the out-of-bounds
array access. If one adds a branch to the example that requires that
the input is no more than $15$ characters, the bug is fixed and CBMC will report that the
proofs of all properties have been successful.

\subsection{Built-In Functions and Types}

The CPROVER framework, which encompasses CBMC, provides built-in functions and types in order
to access internal functionality of the verifier, which
can be used to implement functionality that
the source language itself does not provide.

In addition to the \lstinline{assert(condition)} function provided
by \texttt{assert.h}, there is a
\lstinline{__CPROVER_assert(condition, "description")}
which allows to attach a custom description to properties.

The function \lstinline{__CPROVER_assume(condition)} adds an
expression as a constraint to the program. If the expression evaluates
to false on a path, the execution of this program path aborts without failure. 
Assumptions are used to restrict non-deterministic choices made by the
program. As an example, suppose we wish to model a non-deterministic
choice that returns a number between $1$ and $100$. There is no integer type
with this range. We therefore use \lstinline{__CPROVER_assume} to restrict the
range of a non-deterministically chosen integer:

\begin{lstlisting}
unsigned int nondet_uint();

unsigned int one_to_hundred()
{
  unsigned int result=nondet_uint();
  __CPROVER_assume(result>=1 && result<=100);
  return result;
}
\end{lstlisting}

This function returns the desired integer from 1 to 100. The user must
ensure that the condition given as an assumption is actually
satisfiable by some non-deterministic choice, otherwise the model
checking step will pass vacuously.

Also note that assumptions are never retroactive. They only affect
assertions (or other properties) that follow them in program
order. This is best illustrated with an example. In the following
variant of the above program the assumption has no effect on the assertion, which means
that the assertion will fail:

\begin{lstlisting}
unsigned int nondet_uint();

unsigned int one_to_hundred()
{
  unsigned int result=nondet_uint();
  assert(result<100);
  __CPROVER_assume(result>=1 && result<=100);
  return result;
}
\end{lstlisting}

Assumptions do restrict the search space, but only for assertions that
follow. As an example, this program, with the same assertion now
placed (in program order) after the assumption, will pass:

\begin{lstlisting}
unsigned int nondet_uint();

unsigned int one_to_hundred()
{
  unsigned int result=nondet_uint();
  __CPROVER_assume(result>=1 && result<=100);
  assert(result<100);
  return result;
}
\end{lstlisting}

Beware that non-determinism cannot be used to obtain the effect of
universal quantification in assumptions. For example:

\begin{lstlisting}
int main()
{
  int a[10], x, y;

  x=nondet_int();
  y=nondet_int();
  __CPROVER_assume(x>=0 && x<10 && y>=0 && y<10);

  __CPROVER_assume(a[x]>=0); *@\label{no-quant:l9}@*

  assert(a[y]>=0); *@\label{no-quant:l11}@*
}
\end{lstlisting}

The assertion in Line~\ref{no-quant:l11} fails as \lstinline{x} and
\lstinline{y} need not have the same value.
Line~\ref{no-quant:l9} only ensures that there exists and index
\lstinline{x} such that \lstinline{a[x]>=0}.

\bigskip

\noindent \emph{Built-in Types.}
\lstinline{__CPROVER_bitvector[size]} is used to specify a bit vector
with arbitrary but fixed size. The usual integer type modifiers signed
and unsigned can be applied. The usual arithmetic promotions will be
applied to operands of this type.

\lstinline{__CPROVER_floatbv[total_size][mantissa_size]} specifies
an IEEE-754 floating point number with arbitrary but fixed size.
\lstinline{total_size} is the total size (in bits) of the number, and
\lstinline{mantissa_size} is the size (in bits) of the mantissa, or
significand (not including the hidden bit, thus for single precision
this should be 23).
The IEEE floating-point arithmetic rounding mode can be set by
assigning to the global variable \lstinline{__CPROVER_rounding_mode}.

\lstinline{__CPROVER_fixedbv[total_size][fraction_size]} specifies a
fixed-point bit vector with arbitrary but fixed
size. \lstinline{total_size} is the total size (in bits) of the type,
and \lstinline{fraction_size} is the number of bits after the radix point.

\bigskip

\noindent \emph{Concurrency.}
Asynchronous threads are created by preceding an instruction with a
label with the prefix \lstinline{__CPROVER_ASYNC_}.
Atomic sections are delimited by \lstinline{__CPROVER_atomic_begin()}
and \lstinline{__CPROVER_atomic_end()}.

The complete CPROVER API documentation can be found at
\url{https://www.cprover.org/cprover-manual/api/}.

\subsection{Built-In Library}

Most C programs make use of functions provided by a library. 
Instances
are functions from the standard ANSI-C library such as \lstinline{malloc} or
\lstinline{printf}. The verification of programs that use such functions has two
requirements:
\begin{enumerate}
\item Appropriate header files have to be provided. These header files
  contain declarations of the functions that are to be used.
\item Appropriate definitions have to be provided.
\end{enumerate}

Most C compilers come with header files for the ANSI-C library functions.
CBMC ships definitions of commonly used functions, such as memory
allocation or string manipulation.
These functions often over-approximate the behaviour prescribed by the
C standard to aid sound verification.
An example of such library functions provided by CBMC is (a subset of)
the \texttt{pthread} library, which is used by the following example.


\begin{figure}[!t]
\begin{multicols}{2}
\begin{lstlisting}
#include <assert.h>
#include <pthread.h>

pthread_mutex_t mutex;
int balance = 1000;

void* transaction(void* amount) 
{
  // pthread_mutex_lock(&mutex);

  int current = balance;
  current += *(int *)amount;
  balance = current;

  // pthread_mutex_unlock(&mutex);

  return 0;
}
int main() 
{
  pthread_t t1,t2;
  pthread_mutex_init(&mutex, 0);

  int amount1 = -3000;
  pthread_create(&t1, 0, transaction, &amount1);
  int amount2 = 9000;
  pthread_create(&t2, 0, transaction, &amount2);

  pthread_join(t1, 0);
  pthread_join(t2, 0);
  assert(balance == 6000);

  pthread_mutex_destroy(&mutex);
  return 0;
}
\end{lstlisting}
\end{multicols}
  \captionof{lstlisting}{Source code (\texttt{account.c})}
  \label{prog:concurrency}
\end{figure}

The bank account program in
Listing~\ref{prog:concurrency} uses the \lstinline{pthread} library to
launch two threads to execute transactions on a bank account.
CBMC has support for the \lstinline{pthread} library, so we can
model check this concurrent program by simply running
\texttt{cbmc account.c}.
CBMC reports promptly:
\begin{lstlisting}[language={}]
** Results:
...
account.c function main
[main.assertion.1] line 29 assertion balance == 6000: FAILURE
...
VERIFICATION FAILED
\end{lstlisting}

This program suffers from a race condition.
Race conditions may occur in multi-threaded
programs when the result of a computation depends on the
interleaving of the execution of instructions from
concurrent threads.
By inspecting the trace we can even see why the race condition is
happening:
\begin{lstlisting}[language={}]
...
State 108 file account.c function transaction line 12 thread 1
----------------------------------------------------
  balance=-2000 (11111111 11111111 11111000 00110000)
...
State 137 file account.c function transaction line 12 thread 2
----------------------------------------------------
  balance=9000 (00000000 00000000 00100011 00101000)
...
\end{lstlisting}
Thread 1 withdraws 3000 and sets the balance to -1000, but then thread 2
overwrites the balance with 10000 added to the initial balance,
resulting in 9000 instead of the expected 6000.

Obviously, the program can be repaired by making the update of the
balance atomic. We can uncomment the locks in lines 9 and 13 in
Listing~\ref{prog:concurrency} and verify with CBMC that the program
works correctly now.

\subsection{Test Inputs}

CBMC can be used to automatically generate test inputs that satisfy a
certain code coverage criteria. Common coverage criteria include
branch coverage, condition coverage and Modified Condition/Decision
Coverage (MC/DC). Among others, MC/DC is required by several avionics
software development guidelines to ensure adequate testing of safety
critical software. Briefly, in order to satisfy MC/DC, for every
conditional statement containing Boolean decisions, each Boolean
variable should be evaluated one time to ``true'' and one time to
``false,'' in a way that affects the outcome of the decision.

In the following, we are going to demonstrate how to apply the test
suite generation functionality in CBMC. The program \texttt{pid.c} in
Listing~\ref{prog:pid} is an excerpt from a real-time embedded
benchmark PapaBench~\cite{DBLP:conf/wcet/NemerCSBM06}, and implements
part of a fly-by-wire autopilot for an Unmanned Aerial Vehicle
(UAV). We have adjusted it slightly for our purposes.

\begin{lstlisting}[float=t, firstnumber=28,caption={Part of source code (\texttt{pid.c})}, label={prog:pid}]
void climb_pid_run()
{
  float err=estimator_z_dot-desired_climb;

  float fgaz=CLIMB_PGAIN*(err+CLIMB_IGAIN*climb_sum_err)+
               CLIMB_LEVEL_GAZ+CLIMB_GAZ_OF_CLIMB*desired_climb;

  float pprz=fgaz*MAX_PPRZ;
  desired_gaz=((pprz>=0 && pprz<=MAX_PPRZ) ? pprz : (pprz>MAX_PPRZ ? MAX_PPRZ : 0));

  /** pitch offset for climb */
  float pitch_of_vz=(desired_climb>0) ? desired_climb*CLIMB_PITCH_OF_VZ_PGAIN : 0;
  desired_pitch=NAV_PITCH+pitch_of_vz;

  climb_sum_err=err+climb_sum_err;
  if (climb_sum_err>MAX_CLIMB_SUM_ERR) climb_sum_err=MAX_CLIMB_SUM_ERR;
  if (climb_sum_err<-MAX_CLIMB_SUM_ERR) climb_sum_err=-MAX_CLIMB_SUM_ERR;
}

int main()
{
  while(1)
  {
    /** Non-deterministic input values */
    desired_climb=nondet_float();
    estimator_z_dot=nondet_float();

    /** Range of input values */
    __CPROVER_assume(desired_climb>=-MAX_CLIMB && desired_climb<=MAX_CLIMB);
    __CPROVER_assume(estimator_z_dot>=-MAX_CLIMB && estimator_z_dot<=MAX_CLIMB);

    __CPROVER_input("desired_climb", desired_climb);
    __CPROVER_input("estimator_z_dot", estimator_z_dot);

    climb_pid_run();

    __CPROVER_output("desired_gaz", desired_gaz);
    __CPROVER_output("desired_pitch", desired_pitch);
  }
  return 0;
}
\end{lstlisting}

\begin{figure}
  \centering
  \includegraphics[width=0.5\textwidth]{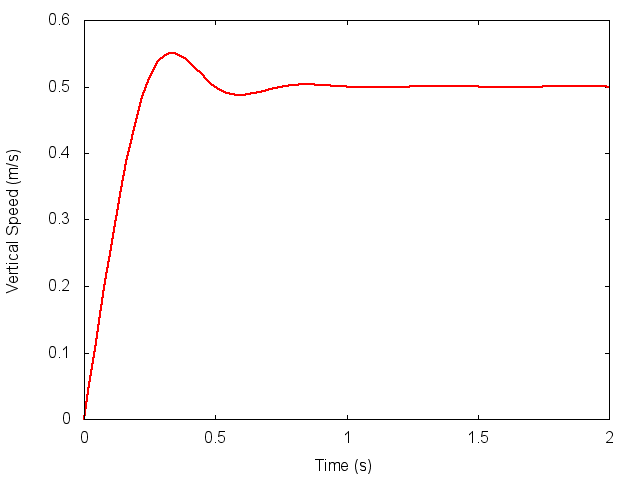}
  \caption{Behaviour of PID controller}
  \label{fig:pid}
\end{figure}

The aim of function \lstinline{climb_pid_run} is to control the vertical
climb of the UAV. It is called from the reactive loop in the
\lstinline{main} function.  The behaviour of this simple controller,
supposing that the desired speed is 0.5 meters per second, is plotted
in Figure~\ref{fig:pid}.

The \lstinline{main} function has been augmented to model
the inputs that are acquired in each time step.
%
%
The functions \lstinline{__CPROVER_input} and
\lstinline{__CPROVER_output} are used to report an input or output
value. Note that they do not generate input or output values, just
report their values. The
first argument is a string constant to distinguish multiple inputs and
outputs (inputs are typically generated using non-determinism, as
described here). The string constant is followed by an arbitrary
number of values of arbitrary types.

Listing~\ref{fig:pid-output} shows a pretty-printed version of a test
suite computing using the following call to CBMC:

\begin{cmd}\label{cmd:pid}
cbmc pid.c -{}-cover mcdc -{}-unwind 6
\end{cmd}

%
\noindent
It shows a test suite that achieves close to 100\% MC/DC
coverage (obviously, the \lstinline{return} statement in \lstinline{main} is not covered).
The test inputs that need to be supplied for each time step are
listed for each test.

\begin{lstlisting}[float=t, language={},caption={Generated test suite for \texttt{pid.c} (Listing~\ref{prog:pid}) using command~\refcmd{cmd:pid}}, label={fig:pid-output}]
Test suite:
Test 1.
  (iteration 1) desired_climb=-1.000000f, estimator_z_dot=1.000000f

Test 2.
  (iteration 1) desired_climb=-1.000000f, estimator_z_dot=1.000000f
  (iteration 2) desired_climb=1.000000f, estimator_z_dot=-1.000000f

Test 3.
  (iteration 1) desired_climb=0.000000f, estimator_z_dot=-1.000000f
  (iteration 2) desired_climb=1.000000f, estimator_z_dot=-1.000000f

Test 4.
  (iteration 1) desired_climb=1.000000f, estimator_z_dot=-1.000000f
  (iteration 2) desired_climb=1.000000f, estimator_z_dot=-1.000000f
  (iteration 3) desired_climb=1.000000f, estimator_z_dot=-1.000000f
  (iteration 4) desired_climb=1.000000f, estimator_z_dot=-1.000000f
  (iteration 5) desired_climb=0.000000f, estimator_z_dot=-1.000000f
  (iteration 6) desired_climb=1.000000f, estimator_z_dot=-1.000000f

Test 5.
  (iteration 1) desired_climb=-1.000000f, estimator_z_dot=1.000000f
  (iteration 2) desired_climb=-1.000000f, estimator_z_dot=1.000000f
  (iteration 3) desired_climb=-1.000000f, estimator_z_dot=1.000000f
  (iteration 4) desired_climb=-1.000000f, estimator_z_dot=1.000000f
  (iteration 5) desired_climb=-1.000000f, estimator_z_dot=1.000000f
  (iteration 6) desired_climb=-1.000000f, estimator_z_dot=1.000000f
\end{lstlisting}

CBMC supports various other coverage criteria apart from MC/DC,
such as branch and condition coverage.
%
%
Moreover, the \lstinline{__CPROVER_cover(condition);} statement can be
used to define a custom coverage criterion.





\section{Verifying Real-World Software} \label{sec:cprover}


Existing software projects usually do not come in a single source file
that may simply be passed to a model checker. Rather, they come in a
multitude of source files in different directories and refer to
external libraries and system-wide options. A build system then
collects the configuration options from the system and compiles the
software according to build rules.

Running software verification
tools on projects like these is greatly simplified by a compiler that
first collects all the necessary models into a single model
file. \texttt{goto-cc} is such a model file extractor.
It uses the compiler's (e.g., GCC's) preprocessor to turn text into actual C
code.
The result of preprocessing is passed on to the
internal C parser (built and evolved as part of the CBMC tools for more than
ten years).  This parser supports several C dialects, including GCC's
extensions, Visual Studio, CodeWarrior, and ARM-CC. Alongside the C dialect
\texttt{goto-cc} also has to (and does) interpret any relevant command line
options of all these tools as they may affect the semantics of the program.
\texttt{goto-cc} builds an intermediate representation, called ``goto
programs'' -- a control-flow graph like representation -- rather than
executable binaries.

Build systems at times first produce executables to use as
part of the build process, or invoke linkers that inspect object
files.
\texttt{goto-cc} can also build hybrid binaries that contain both
executable code as well as models for verification.
To enable this mode, create a link to the \texttt{goto-cc} binary by
the name of \texttt{goto-gcc}.
In this mode, first the original compiler or linker is invoked.
This produces an object file or executable, in ELF format (e.g.,
containing x86/64 bit instructions). 
Next, \texttt{goto-cc} is invoked as either compiler or linker, using the same
command line options as those that were passed to the original compiler or
linker. When compiling, this step, as noted above, produces an intermediate
representation of the compilation unit. To cope with arbitrary build systems,
the resulting intermediate representation is \emph{added as new section} to the
ELF object file or executable. When using \texttt{goto-cc} for linking it thus
reads the extra section from the various input files, performs linking, and then
adds the result of linking onto the output file produced by the original linker.
\texttt{goto-cc} also supports an equivalent approach for OS X, which uses a
different object-file format. There, so-called fat binaries are built to
simulate the described behaviour.
Extensions to support linker scripts in this process are discussed
in~\cite{DBLP:conf/cav/CookKKTTT18}.

Note that adding the intermediate representation onto the original object file
is a key step. The result guarantees that the file remains executable or usable
by the original linker; operations such as renaming or building archives will
always be applied to both the result of unmodified compilation as well as the
intermediate representation, without any extra work being required.
This enables running CBMC on various packages of a Linux
distribution~\cite{DBLP:conf/memics/KroeningT14,DBLP:conf/isola/MalacariaTD16,DBLP:conf/fmcad/CookDKMPPTW20}.
Alternatively, all such steps would need to be traced, e.g., by replacing system
libraries, as is done in
ECLAIR\footnote{\url{https://bugseng.com/products/eclair/discover}}.

Some software projects come with their own libraries. Also, the goal
may be to analyse a library by itself. For this purpose it is possible
to use \texttt{goto-cc} to link multiple model files into a library of model
files. An object file can then be linked against this model
library. For this purpose, \texttt{goto-cc} also features a linker mode.

To enable this linker mode, create a link to the \texttt{goto-cc}
binary by the name of \texttt{goto-ld} (Linux and Mac) or copy the
\texttt{goto-cc} binary to \texttt{goto-link.exe} (Windows). The
\texttt{goto-ld} tool can now be used as a seamless replacement for
the \texttt{ld} tool present on most Unix (-based) systems and for the
link tool on Windows.

Further information can be found at
\url{https://www.cprover.org/cprover-manual/goto-cc/}.

\section{Architecture} \label{sec:architecture}

\begin{figure}[ht]
\centering
\begin{tikzpicture}[line width=.5pt, >=latex,scale=0.8, every node/.style={transform shape}]

\tikzstyle{boxnode} = [rectangle, draw,minimum height=3em, minimum width=7em,
text width=6.5em, align=center, node distance=10em]
\tikzstyle{domainnode} = [rectangle, draw, minimum height = 1.8em, minimum
  width = 8.8em, node distance = 2.5em, align=center]

\node (src) at (0,0) {source program};
\node [boxnode, right of = src] (parse) {C parsing};
\node [boxnode, right of = parse] (typecheck) {Type checking};
\node [boxnode, right of = typecheck] (convert) {GOTO conversion};
\path [->] (src) edge (parse);
\path [->] (parse) edge (typecheck);
\path [->] (typecheck) edge (convert);

\node [boxnode, below of = parse] (lowering) {Analysis \& transformation};
\node [boxnode, below of = typecheck] (instr) {Property instrumentation};
\node [boxnode, below of = convert] (symex) {Symbolic execution};
\draw [->] (convert) |- node [above, xshift = -10em] {GOTO} ($(lowering.north)+(0,0.8)$) -- (lowering);
\path [->] (lowering) edge (instr);
\path [->] (instr) edge (symex);

\node [boxnode, below of = lowering] (encode) {Formula encoding};
\node [boxnode, below of = instr] (solve) {SAT solver};
\draw [->] (symex) |- node [above, xshift = -10em] {SSA} ($(encode.north)+(0,0.8)$) -- (encode);
\node [boxnode, below right of = solve, yshift = 5ex] (cex) {Counterexample};

\node[right = 10em of solve] (true) {
\begin{tikzpicture}
\fill[color=green!50!olive!80](0,.35) -- (.25,0) -- (1,.7) -- (.25,.15) -- cycle;
\end{tikzpicture}
};

\node[right = 3em of cex] (false) {
\begin{tikzpicture}
\node[scale=2, red] {\ding{55}};
\end{tikzpicture}
};

\path [->] (encode) edge (solve);
\path [->] (solve) edge (true);
\path [->] (solve) edge (cex);
\path [->] (cex) edge (false);

\node [right = 4em of convert] (front) {Front End};
\node [below of = front, node distance = 11em] (middle) {Middle End};
\node [below of = middle, node distance = 12em] (back) {Back End};

\end{tikzpicture}
\caption{CBMC Overview
}
\label{fig:arch}
\end{figure}
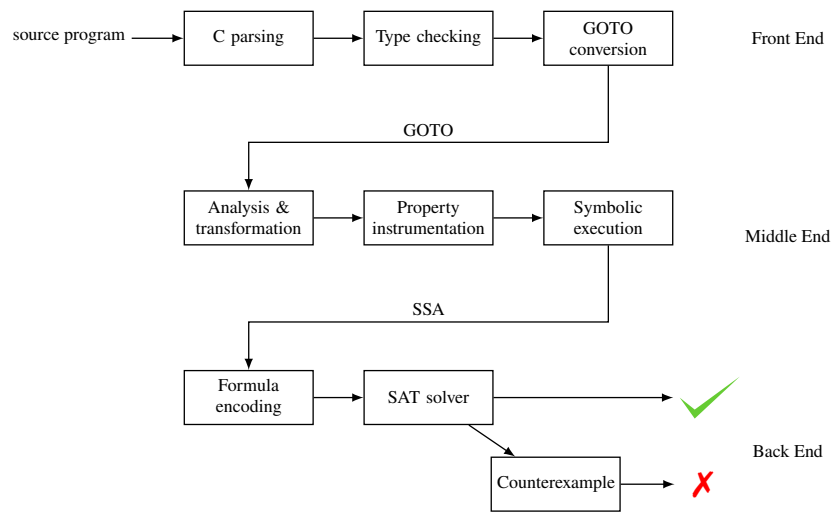

Bounded model checkers such as CBMC reduce questions about program
paths-to constraints that can be solved by off-the-shelf SAT or SMT
solvers. With the SAT back end, and given a program annotated with
assertions, CBMC outputs a CNF formula the solutions of which describe
program paths leading to assertion violations. In order to do so, CBMC
performs the following main steps, which are outlined in
Figure~\ref{fig:arch}, and are explained below.

\subsection{Command-line Front End}

The command line front end processes options given by the user and
configures CBMC accordingly.
The general command syntax to call CBMC is

\begin{cmd}
  cbmc [options ...] [file.c ...]
\end{cmd}  

\noindent
where {\tt [options ...]} are described below and {\tt [file.c ...]}
are zero or more source file names.
Typical user-supplied parameters are loop unwinding limits, or the
system architecture to be assumed, i.e., the bit-width of basic data
types and endianness, and operating system specific configuration
parameters.
If the user chooses not to override the platform configuration, then
the system architecture configuration defaults to the specification of
the platform CBMC has been compiled on.
In the following we provide a description of the most commonly used
command-line options; further options controlling specific parts of
CBMC's architecture are discussed in subsequent sections.

\begin{itemize}
\item General options:
  
  \begin{itemize}
    \item {\tt -{}-help}, {\tt -h}, {\tt -?}: Display copyright
      information and the list of command line options.
    \item {\tt -{}-version}: Show the current version.
  \end{itemize}

\item Platform configuration:
  \begin{itemize}
    \item {\tt -{}-function $f$}: Use function $f$ as program entry
      point instead of ``\lstinline{main}''.
    \item {\tt -I $path$}: Add $path$ to the C preprocessor's search
      path for expanding {\tt \#include} directives.
      This option may be given multiple times, which is the case for
      {\tt -D $macro$} as well:
    \item {\tt -D $macro$}: Define the C preprocessor macro $macro$,
      where $macro$ is either only a macro name or of the form {\tt
      $name$=$value$}.
    \item {\tt -{}-16}, {\tt -{}-32}, {\tt -{}-64}: Set the bit-width of
      the C type \lstinline{int} to 16, 32, or 64 bits, respectively.
    \item {\tt -{}-LP64}, {\tt -{}-ILP64}, {\tt -{}-LLP64}, {\tt
      -{}-ILP32}, {\tt -{}-LP32}: Set the bit-widths of int, long, and
      pointers as defined in~\cite{UNIX98:LP64}.
    \item {\tt -{}-little-endian}, {\tt -{}-big-endian}: Set endianness
      for conversions between words and bytes.
    \item {\tt -{}-unsigned-char}: Make \lstinline{char} type unsigned.
    \item {\tt -{}-ppc-macos}, {\tt -{}-i386-macos}, {\tt
      -{}-i386-linux}, {\tt -{}-i386-win32}, {\tt -{}-win32}, and {\tt
      -{}-winx64}: Set platform-specific defines, bit-widths, and
      endianness according to the given processor and operation-system
      combination.
  \end{itemize}

\item Options controlling the user interface:
  \begin{itemize}
    \item {\tt -{}-json-ui}, {\tt -{}-xml-ui}: Change CBMC's output to
      JSON or XML formatted text, respectively. {\tt
      -{}-json-interface}, {\tt -{}-xml-interface} extend this to
      consuming options via JSON or XML, respectively.
      Such output (and input) is more suitable for machine processing.
    \item {\tt -{}-verbosity $n$}: Sets the amount of status
      information printed while running CBMC.
      $n = 0$ disables all output other than test cases, $n \geq 1$
      enables error messages, $n \geq 2$ adds warnings, $n \geq 6$
      enables progress information ($n = 6$ is the default), $n \geq
      8$ adds statistics, and $n \geq  9$ yields debugging
      information.
  \end{itemize}
\end{itemize}

\subsection{Language Front End} \label{language-front-end}

CBMC uses ``GOTO programs'' as intermediate representation.
In this language, all non-linear control flow, such as
if/switch-statements, loops and jumps, is translated to equivalent
\emph{guarded goto} statements.
These statements are gotos that include optional guards, such that
these guarded gotos also suffice to model if/else branching.
The most important classes of statements left at this intermediate
level are assignments, gotos, function calls, declarations,
assertions, and assumptions.

For C/C++ input source, CBMC arrives at this intermediate
representation by first invoking a C preprocessor (for example, {\tt
cl} on Microsoft Windows systems or {\tt gcc -E} on Unix-like systems,
but other compiler tool-chains are supported as well as discussed in
Section~\ref{sec:cprover}) and then passing
the result to CBMC's C or C++ parser.
The output of the preprocessor can inspected by calling CBMC with {\tt
-{}-preprocess}.
CBMC uses its own C and C++ parser rather than relying on existing
tool chains, which enables supporting multiple C/C++ dialects
including various GCC extensions.
The C or C++ parser yields a parse tree annotated with source file-
and line information.
The parse tree with annotated type information can be inspected by
calling CBMC with {\tt -{}-show-parse-tree}.
Other language front ends, for example the Java
front end~\cite{CKKST18}, have a different parsing work-flow, but
ultimately also proceed to type checking:

Type checking populates a symbol table by traversing the parse tree,
collecting all type names and symbol identifiers, and assigning
bit-level type information to each symbol and expression that is
found.
To view the symbol table, invoke CBMC with {\tt
-{}-show-symbol-table}.
CBMC aborts if any inconsistencies are detected by type checking.
As an experiment, comparing the output of {\tt cbmc -{}-16 abs.c}
vs.~{\tt cbmc -{}-32 abs.c} shows how type checking affects
bit-level types, and thus also the expansion of constants to different
bit vectors:
\begin{figure}[!h]
\columnsep-6em
\begin{multicols}{2}
\begin{lstlisting}[numbers=none]
constant
 * type: signedbv
     * width: 16
     * #c_type: signed_int
 * value: 0000000000000000
 * #source_location:
   * file: abs.c
   * line: 4
   * function: abs
   * working_directory: /home
 * #base: 10

constant
 * type: signedbv
     * width: 32
     * #c_type: signed_int
 * value: 00000000000000000000000000000000
 * #source_location:
   * file: abs.c
   * line: 4
   * function: abs
   * working_directory: /home
 * #base: 10
\end{lstlisting}
\end{multicols}
\end{figure}

When type checking succeeds, CBMC generates one GOTO program for each
procedure or method found in the parse tree.
Furthermore, it adds a new main function that first calls an
initialisation function for objects with static lifetime and then
calls the original program entry function.

For the earlier example of \texttt{abs.c} as shown in
Listing~\ref{prog:abs}, running

\begin{cmd}\label{cmd:show-goto}
  cbmc -{}-function abs -{}-signed-overflow-check -{}-show-goto-functions
\end{cmd}

\noindent
yields the output shown in Listing~\ref{show-goto-functions}.
\begin{figure}[t]
\columnsep6em
\begin{multicols}{2}
\begin{lstlisting}[]
abs *@\label{gf:l1}@*
   // 0 file abs.c line 2 function abs *@\label{gf:l2}@*
   signed int y; *@\label{gf:l3}@*
   // 1 file abs.c line 2 function abs *@\label{gf:l4}@*
   y = x; *@\label{gf:l5}@*
   // 2 file abs.c line 4 function abs *@\label{gf:l6}@*
   IF !(x < 0) THEN GOTO 1 *@\label{gf:l7}@*
   // 3 file abs.c line 5 function abs *@\label{gf:l8}@*
   ASSERT !(x == -2147483648) *@\label{gf:l9}@*
   // 4 file abs.c line 5 function abs *@\label{gf:l10}@*
   y = -x; *@\label{gf:l11}@*
   // 5 file abs.c line 8 function abs *@\label{gf:l12}@*
1: abs#return_value = y; *@\label{gf:l13}@*
   // 6 file abs.c line 8 function abs *@\label{gf:l14}@*
   dead y; *@\label{gf:l15}@*
   // 7 file abs.c line 9 function abs *@\label{gf:l16}@*
   END_FUNCTION *@\label{gf:l17}@*

__CPROVER_initialize *@\label{gf:l19}@*
   // 8 file <built-in-additions> line 20 *@\label{gf:l20}@*
   // Labels: __CPROVER_HIDE *@\label{gf:l21}@*
   __CPROVER_rounding_mode = 0; *@\label{gf:l22}@*
   // 9 no location *@\label{gf:l23}@*
   END_FUNCTION *@\label{gf:l24}@*

__CPROVER__start *@\label{gf:l26}@*
   // 10 no location *@\label{gf:l27}@*
   __CPROVER_initialize(); *@\label{gf:l28}@*
   // 11 file abs.c line 1 *@\label{gf:l29}@*
   signed int x; *@\label{gf:l30}@*
   // 12 file abs.c line 1 *@\label{gf:l31}@*
   x = NONDET(signed int); *@\label{gf:l32}@*
   // 13 file abs.c line 1 *@\label{gf:l33}@*
   INPUT("x", x); *@\label{gf:l34}@*
   // 14 file abs.c line 1 *@\label{gf:l35}@*
   abs(x); *@\label{gf:l36}@*
   // 15 file abs.c line 1 *@\label{gf:l37}@*
   return$'$ = abs#return_value; *@\label{gf:l38}@*
   // 16 file abs.c line 1 *@\label{gf:l39}@*
   dead abs#return_value; *@\label{gf:l40}@*
   // 17 file abs.c line 1 *@\label{gf:l41}@*
   OUTPUT("return", return$'$); *@\label{gf:l42}@*
   // 18 no location *@\label{gf:l43}@*
   dead x; *@\label{gf:l44}@*
   // 19 no location *@\label{gf:l45}@*
   END_FUNCTION *@\label{gf:l46}@*
\end{lstlisting}
\end{multicols}
\begin{lstlisting}[caption={GOTO programs output by command~\refcmd{cmd:show-goto} for \texttt{abs.c} (Listing~\ref{prog:abs})},label={show-goto-functions}]
\end{lstlisting}
\end{figure}
In this output, each GOTO program (each procedure) starts with its
name (lines~\ref{gf:l1}, \ref{gf:l19}, and \ref{gf:l26}) and ends with
an {\tt END\_FUNCTION} GOTO-program instruction (lines~\ref{gf:l17},
\ref{gf:l24}, and \ref{gf:l46}).
Each instruction includes the source location that it originates from
(lines~\ref{gf:l2}, \ref{gf:l4}, etc.).
The instruction itself is printed in a C-like syntax.
For example, lines~\ref{gf:l3} and \ref{gf:l30} denote declarations of
variables, which may go out of scope -- each such point is denoted by
a corresponding {\tt dead} instruction, as can be found in
lines~\ref{gf:l15} and \ref{gf:l44}.
Instructions in lines~\ref{gf:l5}, \ref{gf:l13}, among others, denote
assignments.
Furthermore line~\ref{gf:l13} carries a label: it is the branch target
of the (conditional) \lstinline+goto+ in line~\ref{gf:l7}.
This conditional \lstinline+goto+ encodes the control flow resulting
from the \lstinline+if+ statement in line~\ref{abs-out:l4} of
Listing~\ref{prog:abs}.

\bigskip

\noindent
\emph{Analysis and Transformation.}
The case of line~\ref{gf:l13} is peculiar in that it constitutes the
result of return-instruction removal: to keep the classes of
instructions to be considered by analyses as small as possible, we
simulate the effect of \lstinline+return+ statements via
\lstinline+goto+ and assignments to global, thread-local variables.
Further transformations, though not applicable to this example,
include replacing function pointers.
First, function pointers are resolved via a light-weight static
analysis that checks for type compatibility between formal parameters
of declared functions and the actual parameters at the point of call
through a function pointer.
All matching targets are combined to a list of conditional calls,
where a branch is taken if the actual value of the function pointer
matches the address of the target function. Thereby we arrive at a
static call graph.

\bigskip

\noindent
\emph{Property Instrumentation.}
In line~\ref{gf:l9} we find the generated assertion as we used {\tt
-{}-signed-overflow-check}.
This assertion guards the possible undefined behaviour resulting from
negation in line~\ref{gf:l11}.
Such assertions are generated using light-weight and
over-approximating data-flow analyses as discussed in
Section~\ref{sec:usecases}.

Options controlling this instrumentation process include:
\begin{itemize}
  \item {\tt -{}-bounds-check}: Ensure each indexed access to an array
    is within its bounds.
  \item {\tt -{}-pointer-check}: Ensure deferencing is only
    used with pointers to live objects.
  \item {\tt -{}-div-by-zero-check}: Ensure that no divisor is zero.
  \item {\tt -{}-signed-overflow-check}, {\tt -{}-unsigned-overflow-check}: Check the absence of
    arithmetic over- and underflow on signed or unsigned integers,
    respectively.
  \item {\tt -{}-conversion-check}: Ensure that type casts are not
    applied to values that could not be represented in the target
    type.
  \item {\tt -{}-undefined-shift-check}: Ensure shifts do not exceed
    the bit-width of the type of the object being shifted.
  \item {\tt -{}-float-overflow-check}, {\tt -{}-nan-check}: Ensure that floating-point
    operations do not result in positive or negative infinity, or
    not-a-number, respectively.
  \item {\tt -{}-enum-range-check}: Ensure that enum-typed objects
    never take a value other than the declared enum constants for that
    type.
\end{itemize}

\bigskip

\noindent
\emph{Using GOTO Programs.}
GOTO programs as described above can be serialised and deserialised
and and from a custom binary format.
This approach enables the use of \texttt{goto-cc} as discussed in
Section~\ref{sec:cprover}.
Further tools that work on GOTO programs include
\texttt{goto-instrument}, a transformation tool, and
\texttt{goto-diff}, implementing the equivalent of the text-based Unix
tool \texttt{diff} for GOTO programs.

\subsection{Symbolic Execution}


Options controlling program instrumentation and loop unwinding:
  \begin{itemize}
    \item {\tt -{}-no-library}: By default CBMC ships an abstracted
      version of system library functions.
      This options disables inclusion of such code.
    \item {\tt -{}-show-goto-functions}: Display the GOTO functions
      after instrumentation as described in
      Section~\ref{language-front-end}.
      This option is primarily useful for debugging purposes.
    \item {\tt -{}-no-assumptions}: The programmer can add assumptions
      to the program under scrutiny using the
      \lstinline{__CPROVER_assume($x$)} built-in.
      That is, for all paths considered by CBMC, the property~$x$ must
      evaluate to true true at the program point where the built-in was
      inserted.
      If {\tt -{}-no-assumptions} is set, assumptions will be ignored.
    \item {\tt -{}-function $f$}: Use function $f$ as program entry
      point instead of ``\lstinline{main}''.
    \item {\tt -{}-depth $k$}: Perform at most~$k$ steps along any path
      while symbolically executing the program.
      This results in unsound verification, unless~$k$ steps suffice
      to reach all states.
    \item {\tt -{}-unwind $k$}, {\tt -{}-unwindset $L$:$k$,...}, {\tt
      -{}-show-loop-ids}: Unwind loops, recursions, and backward
      gotos at most $k$ times.
      With {\tt -{}-unwindset $L$:$k$,...} the unwinding bound $k$ is
      set for loop with id $L$ only, where $L$ can be found using {\tt
      -{}-show-loop-ids}, which lists all loops with their identifiers.

      Successful verification while using {\tt -{}-unwind} or {\tt
      -{}-unwindset} is unsound, unless the specified bounds amount to complete loop
      unwinding.
      To ensure sound (but possibly incomplete) verification, add
      {\tt -{}-unwinding-assertions}:

    \item {\tt -{}-unwinding-assertions}, {\tt -{}-partial-loops}:
      Whenever the loop unwinding bounds specified using {\tt
      -{}-unwind} or {\tt -{}-unwindset} are reached, CBMC inserts an
      assumptions that the loop condition indeed no longer holds, i.e.,
      the loop would indeed be left.
      This may rule out feasible execution paths, and thus results in
      unsound verification as noted above.
      To avoid this source of unsoundness, {\tt
      -{}-unwinding-assertions} can be specified such that instead of
      assumptions assertions are inserted.
      The assertion checks that the loop exit condition is indeed always
      fulfilled, i.e., the number of unwinding steps was sufficient.

      With the parameter {\tt -{}-partial-loops}, neither assumptions
      nor assertions are generated during loop unwinding.
      This may be useful for experiments, but does result in
      verification of a possibly very different program, because each
      loop is replaced by a fixed number of conditional repetitions of
      the loop body without any checks whether the loop condition
      evaluates to false at the end.
      This may result in traces that cannot occur in the original
      program.
  \end{itemize}

As CBMC implements a variant of bounded model checking it has to pay
special attention to loops.
Unlike the original bounded model checking algorithm presented
in~\cite{BCCZ99},
CBMC currently does not increase the maximum length of paths as bounded model
checking proceeds, and is thus not complete.
Instead, CBMC eagerly unwinds loops up to a fixed bound, which can be specified
by the user on
a per-loop basis or globally, for all loops.
In the course of this unwinding step, CBMC also translates the GOTO functions
to static single assignment (SSA)
form~\cite{DBLP:conf/popl/AlpernWZ88,DBLP:conf/popl/RosenWZ88,DBLP:conf/popl/CytronFRWZ89}.
At the end of this process the program is represented as a mathematical
equation over renamed program variables in guarded statements.
The guards determine whether, given a concrete program execution, an assignment
is actually made.

\pscmt{Explain characteristics/limitations/advantages of BMC}

The basic idea of CBMC is to model a program's execution up to a
bounded number of steps. Technically, this is achieved by a process
that essentially amounts to unwinding loops. This concept is best
illustrated with a generic example:

\begin{lstlisting}
int main(int argc, char **argv) {
  while(cond) {
    BODY CODE
  }
}
\end{lstlisting}

A BMC instance that will find bugs with up to five iterations of the
loop would contain five copies of the loop body, and essentially
corresponds to checking the following loop-free program:

\begin{lstlisting}
int main(int argc, char **argv) {
  if(cond) {
    BODY CODE COPY 1
    if(cond) {
      BODY CODE COPY 2
      if(cond) {
        BODY CODE COPY 3
        if(cond) {
          BODY CODE COPY 4
          if(cond) {
            BODY CODE COPY 5
          }
        }
      }
    }
  }
}
\end{lstlisting}

Note the use of the if statement to prevent the execution of the loop
body in the case that the loop ends before five iterations are
executed. The construction above is meant to produce a program that is
trace equivalent with the original programs for those traces that
contain up to five iterations of the loop.

In many cases, CBMC is able to determine automatically an upper bound
on the number of loop iterations. This may even work when the number
of loop unwindings is not constant. Consider the following example:

\begin{lstlisting}
_Bool f();

int main()
{
  for(int i=0; i<100; i++)
  {
    if(f()) break;
  }
  assert(0);
}
\end{lstlisting}

The loop in the program above has an obvious upper bound on the number
of iterations, but note that the loop may abort prematurely depending
on the value that is returned by \texttt{f()}. CBMC is nevertheless
able to automatically unwind the loop to completion.

This automatic detection of the unwinding bound may fail if the number
of loop iterations is highly data-dependent. Furthermore, the number
of iterations that are executed by any given loop may be too large or
may simply be unbounded. For this case, CBMC offers the command-line
option \texttt{-{}-unwind} $B$, where $B$ denotes a number that
corresponds to the maximal number of loop unwindings CBMC performs on
any loop.

Note that the number of unwindings is measured by counting the number
of backjumps. In the example above, note that the condition
\lstinline{i<100} is in fact evaluated 101 times before the loop
terminates. Thus, the loop requires a limit of 101, and not~100.

In \cite{DBLP:conf/cav/AlglaveKT13}
we presented an extension to perform efficient bounded model checking
of concurrent programs, which symbolically encodes partial orders over
read and write accesses to shared variables.

\pscmt{more details on concurrency?}

\subsection{SAT/SMT Back Ends}

The resulting equation is translated into a CNF formula by bit-precise
modelling of all expressions plus the Boolean guards
(cf.~\cite{DBLP:conf/dac/ClarkeKY03}).  Here it should be noted that
CBMC also supports other decisions procedures as back ends, such as
SMT (satisfiability modulo theories)
solvers~\cite{DBLP:conf/lpar/NieuwenhuisOT04,DBLP:journals/jacm/NieuwenhuisOT06},
in which case an encoding other than CNF is used.
These back ends can be selected using command-line options such as
{\tt -{}-smt2} (to use the default SMT2 solver, currently Z3), {\tt -{}-z3} to use
Z3~\cite{DBLP:conf/tacas/MouraB08}, or {\tt -{}-cvc4} to select
CVC4~\cite{DBLP:conf/cav/BarrettCDHJKRT11}.

The CNF formula, which can be printed using {\tt -{}-dimacs}, is passed
to the SAT solver, which tries to find a satisfying assignment.  Here,
such an assignment corresponds to a path
violating at least one of the assertions in the program under
scrutiny.  Conversely, if the formula is unsatisfiable, no assertion
can be violated \emph{with the given unwinding bounds}.
CBMC supports multiple properties in the program and
queries the solver iteratively in order to decide the
result for each of the properties.

If a satisfying assignment was found, the bounded model checker has
determined a counterexample to the specification given in terms of
assertions.  To turn the model of the SAT formula into information
useful for the user of CBMC, it is translated into a list of
assignments.  CBMC finds this sequence by consulting the equation of
guarded statements: each statement with a guard evaluating to true
under the computed model constitutes an assignment occurring in the
counterexample.  The actual values being assigned are also found in
the model of the SAT formula.
The resulting counterexample output is as previously shown in
Listing~\ref{output:abs3}.
    
\section{Performance History}\label{sec:history}

CBMC has been continuously developed for more than 15 years,
but has it actually become better over the years?
To answer this question, we benchmarked 22 CBMC versions from version
4.0 (June 2011) to 5.20 (December 2020).%
\footnote{Ports of the older versions are available in the \texttt{cbmc-x.y-patch} branches in \url{https://github.com/diffblue/cbmc}. The corresponding SV-COMP wrapper scripts are in the \texttt{cbmc-x.y} branches in \url{https://github.com/diffblue/cprover-sv-comp}.}
We used 9 categories from SV-COMP%
\footnote{ReachSafety-Arrays, ReachSafety-BitVectors, ReachSafety-ControlFlow, ReachSafety-Floats, ReachSafety-Heap, ReachSafety-Loops, MemSafety-Arrays, MemSafety-Heap, MemSafety-LinkedLists; version \url{https://github.com/sosy-lab/sv-benchmarks/tree/b8369a395d4749eb7eee1c3bd8149a3dc799e7f3}}
and ran them using BenchExec%
\footnote{\url{https://github.com/sosy-lab/benchexec/releases/tag/1.17}}
on a machine with Ubuntu 16.04 and an Intel Xeon Platinum 8175M CPU,
2.50GHz with resource limits of 15\,GB and 900\,s.

\begin{figure}[t]
  \includegraphics[width=\textwidth]{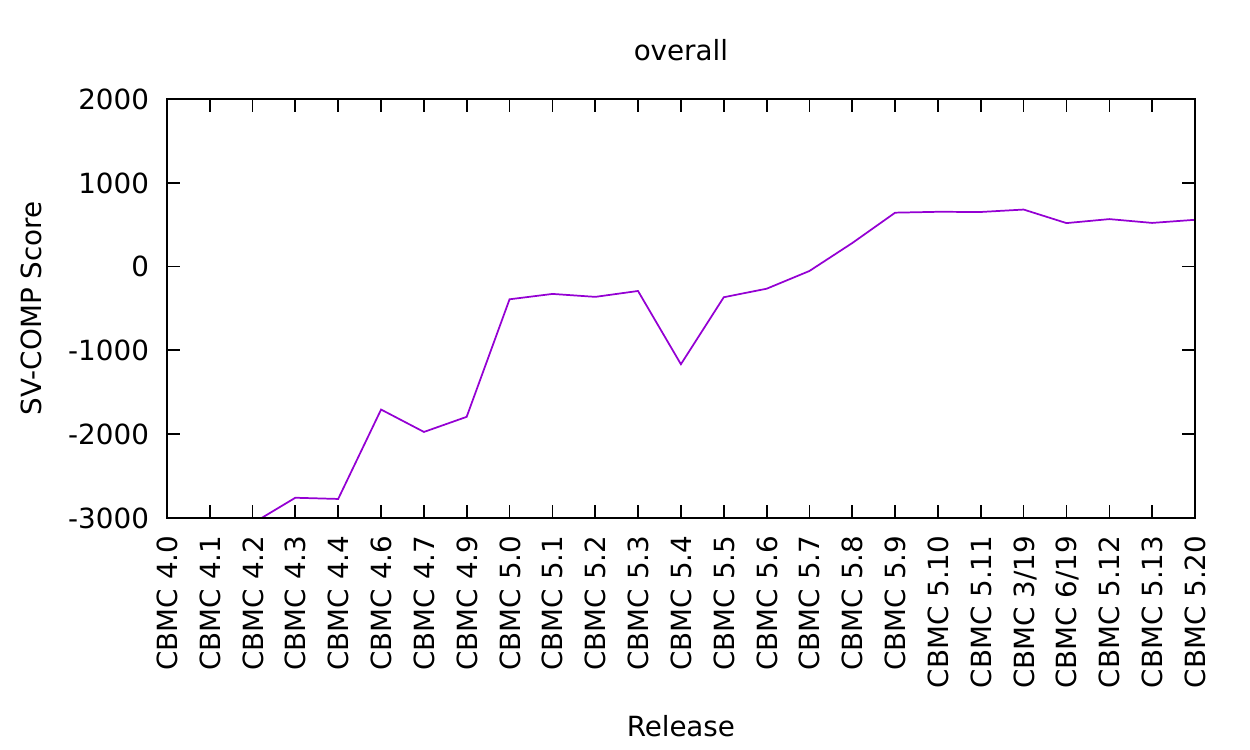}
  \caption{Score time line over all selected benchmarks
  }
  \label{fig:score-timeline-overall}
\end{figure}

\begin{figure}[t]
  \includegraphics[width=\textwidth]{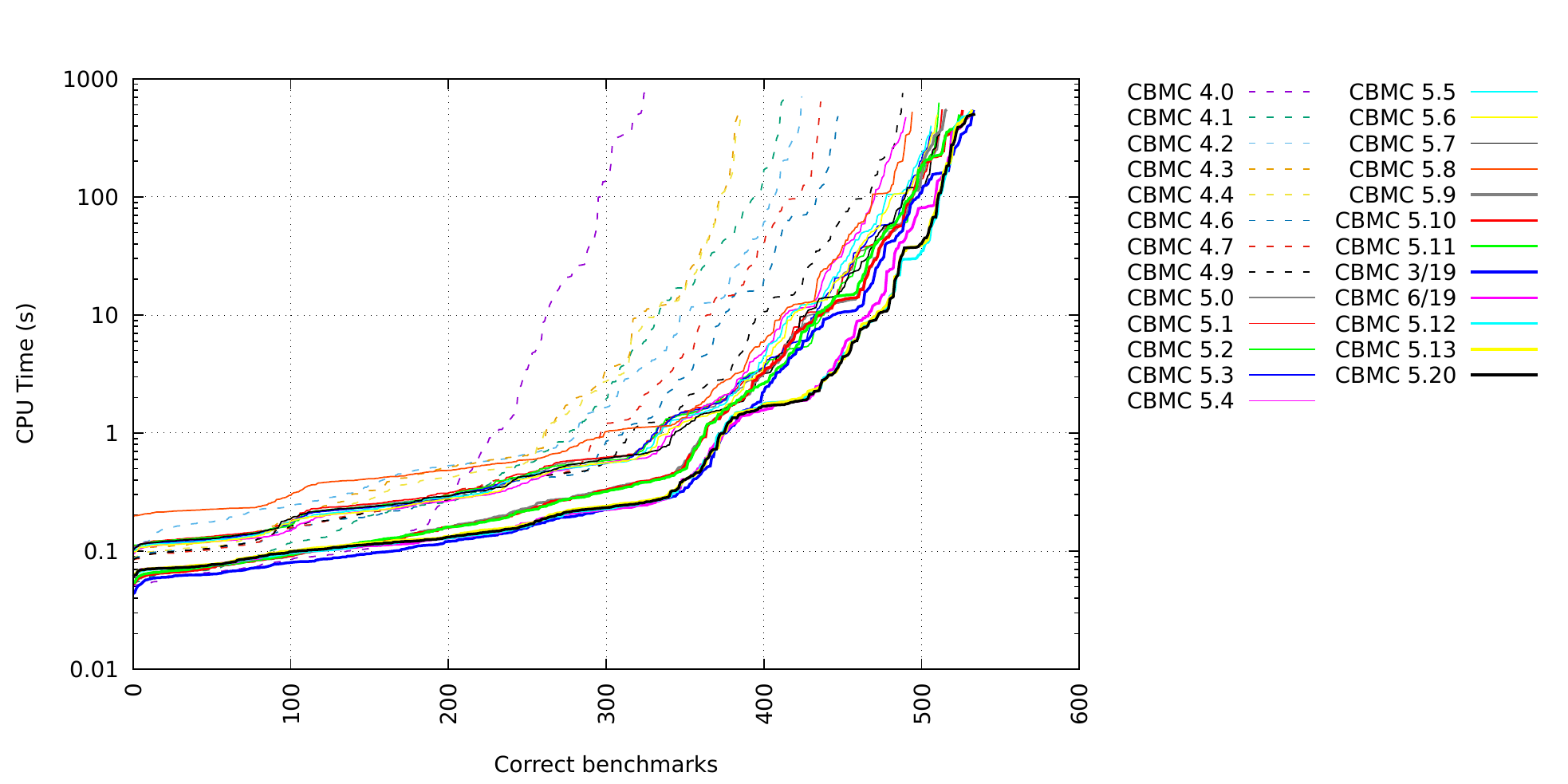}
  \caption{CPU time quantile plot over all selected benchmarks
  }
  \label{fig:cpu-cactus-overall}
\end{figure}

\begin{figure}[t]
  \includegraphics[width=\textwidth]{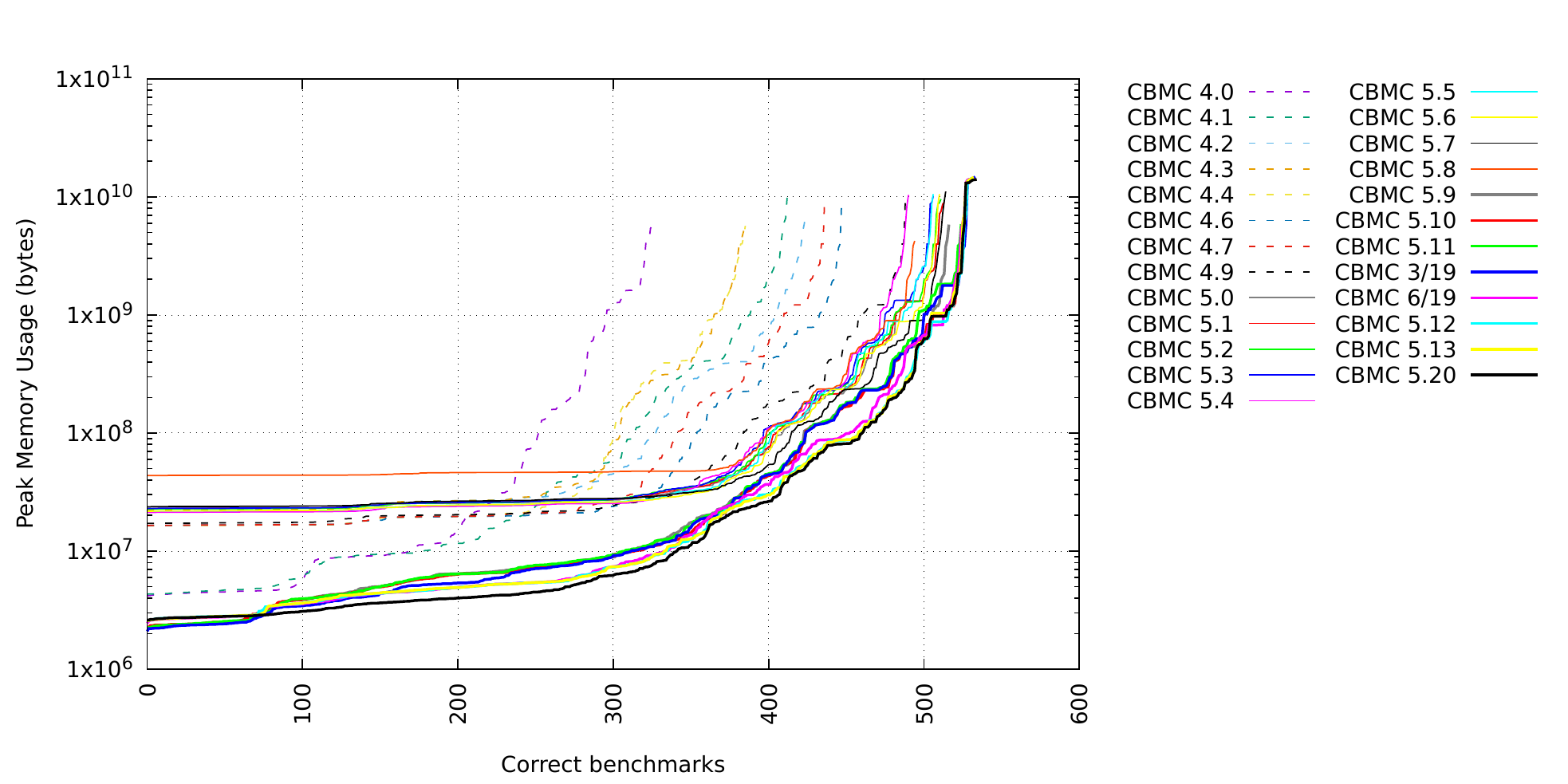}
  \caption{Peak memory usage quantile plot over all selected benchmarks
  }
  \label{fig:mem-cactus-overall}
\end{figure}

Figure~\ref{fig:score-timeline-overall} shows the evolution of
CBMC's SV-COMP score%
\footnote{One point for finding a bug in an unsafe benchmark; -32~points
  for incorrectly claiming it safe. 2~points for proving a safe
  benchmark correct; -16~points for incorrectly reporting a bug.}
on the all the selected benchmarks.
Note that the scores cannot be compared with the official SV-COMP results
because the rules changed over the years as well as the benchmark sets.
CBMC won SV-COMP~2014 with a version based on 4.5, and was ranked
third in SV-COMP~2015 with a version based on 4.9.

In terms of SV-COMP score, CBMC has improved substantially in the more than
9~years spanned by these versions. In particular, the versions towards
CBMC~5.0 (2014-2015) were a huge improvement in comparison to early
CBMC~4.x versions (2011-2013).
A second wave of improvements is visible from CBMC~5.7 to 5.9 (2017-2018).


Looking at the quantile plot of CPU time in
Figure~\ref{fig:cpu-cactus-overall}, we can also observe these
improvements up to CBMC~5.0 (with a notable jump from 4.4 to 4.6 and
4.6 to 5.0) followed by some stagnation up to CBMC~5.7 (and a
regression in~5.4).
%
%
There was, however, substantial speed up from CBMC~5.7 to 5.9 (with a regression
in 5.8) and further less noteworthy speed improvements in most
recent versions.
%
%
The big improvements from CBMC~4.4 to 4.6, 4.6 to 5.0 and 5.7 to 5.9
also brought about a major reduction in memory consumption, as the
quantile plot of peak memory usage in
Figure~\ref{fig:mem-cactus-overall} shows.
%
These improvements can mainly be attributed to enhanced
expression simplification before encoding SSA into a SAT formula.
%
CBMC~5.9 introduced on-demand definitions for compiler-built-in
functions, which gave a constant speed-up of half a second. It is visible
on benchmarks with short runtime.

\begin{figure}[t]
  \includegraphics[width=\textwidth]{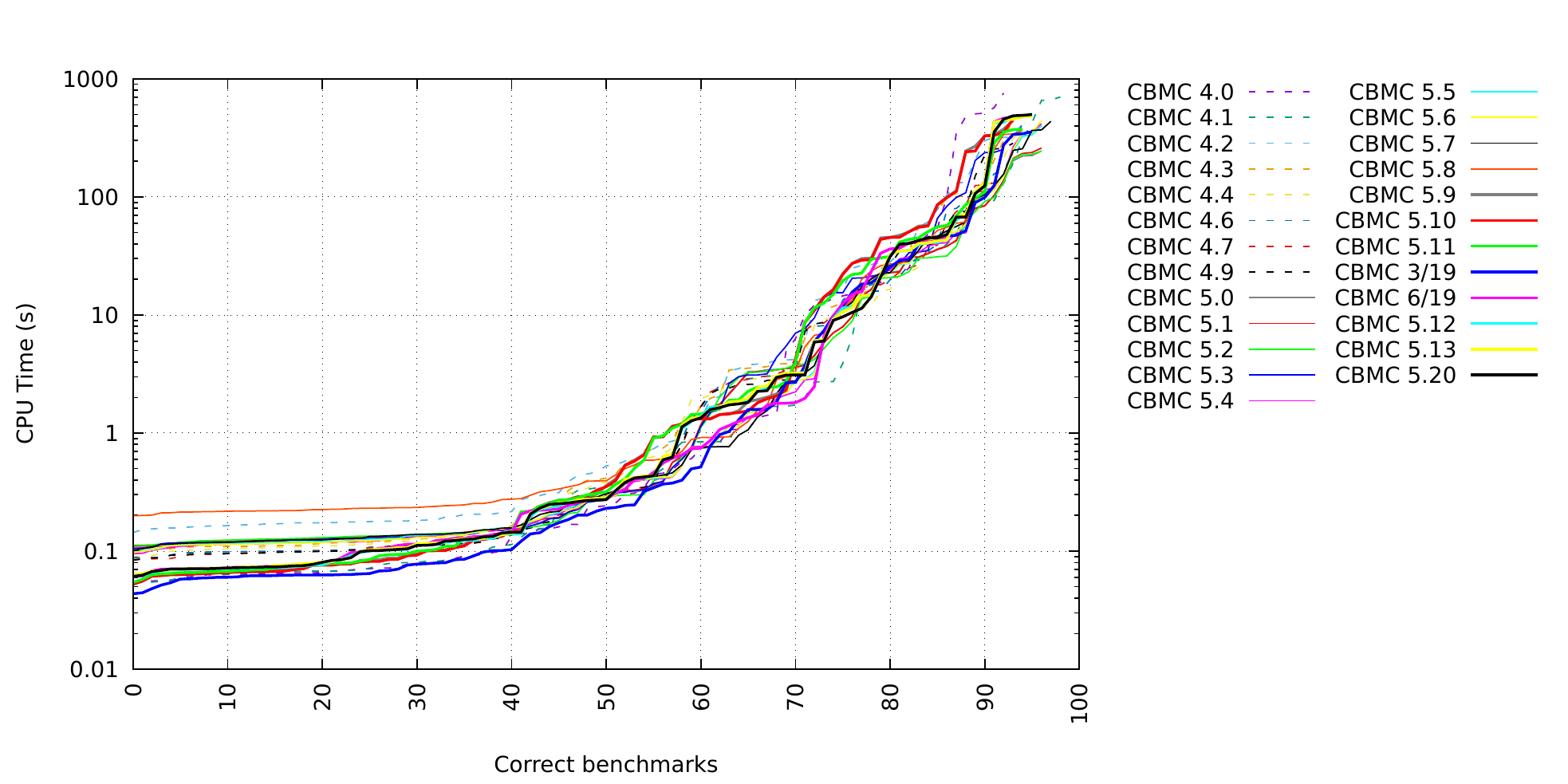}
  \caption{CPU time quantile plot over ReachSafety-Loops
  }
  \label{fig:cpu-cactus-ReachSafety-Loops}
\end{figure}

The improvements were not uniform over all the categories.  For
example, CBMC's performance on the ReachSafety-Loops category shows
only small variations (see
Figure~\ref{fig:cpu-cactus-ReachSafety-Loops}). These benchmarks are
quite simple integer programs, which already CBMC~4.0 supported very
efficiently and solved roughly the same number of benchmarks at a
comparable speed as the latest version.

\begin{figure}[t]
  \includegraphics[width=\textwidth]{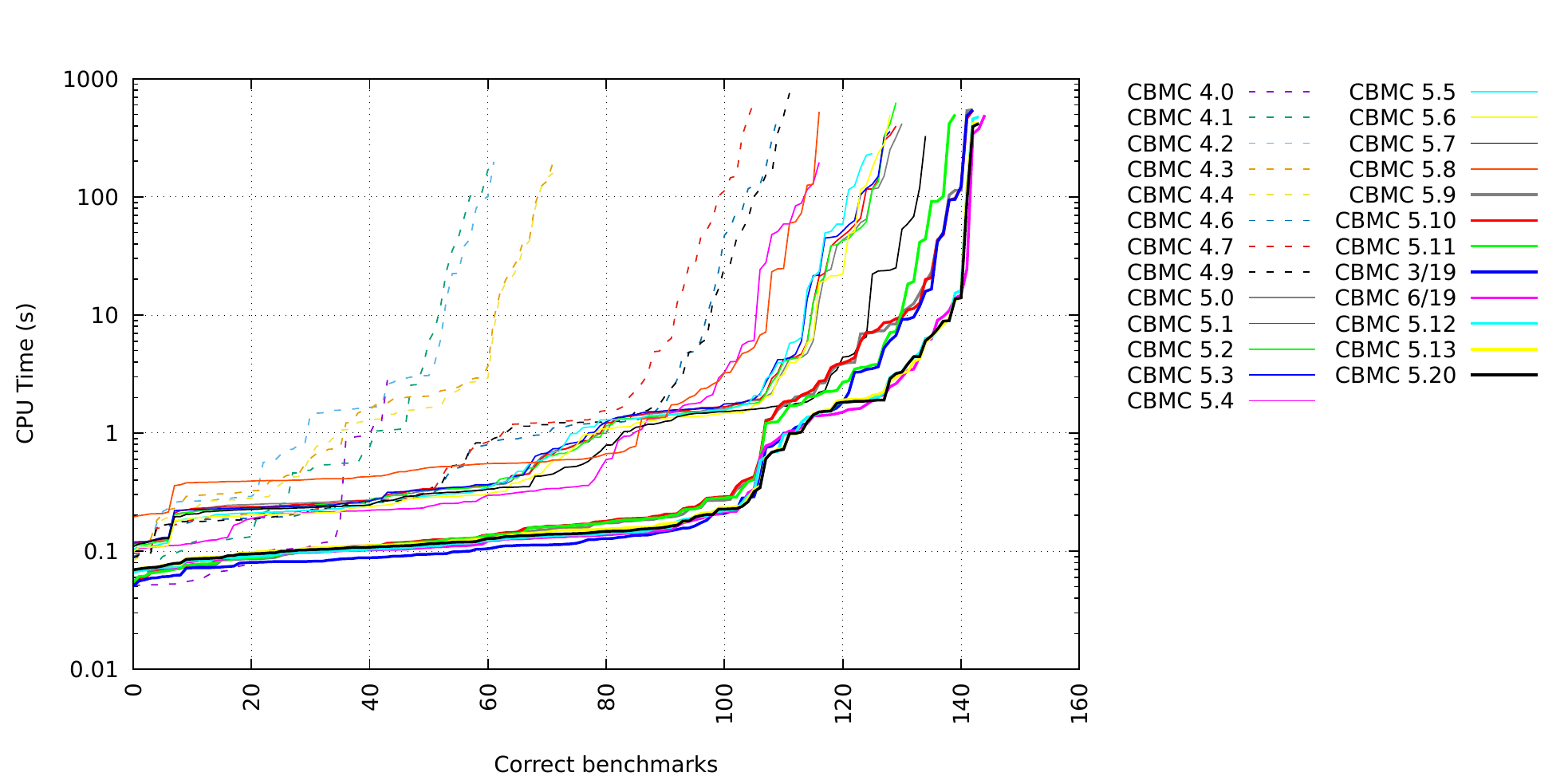}
  \caption{CPU time quantile plot over MemSafety-Heap
  }
  \label{fig:cpu-cactus-MemSafety-Heap}
\end{figure}

A totally different picture can be seen in Figure~\ref{fig:cpu-cactus-MemSafety-Heap}
for the MemSafety-Heap benchmarks. These benchmarks use more complex
language features such as dynamic memory allocation, pointers and structs.
CBMC has not only become significantly faster, but also produces far
fewer incorrect results (from a sixth incorrect results in CBMC~4.6 down to
a single incorrect result since CBMC~5.9).
The gap between CBMC~4.4 and 4.6 is due to the introduction of a
memory leak instrumentation, which enabled proving these properties.

The latest improvements between 3/19 and 5.20 were due to significant
enhancements in the symbolic execution.
For example, the data structures
for performing constant propagation and storing points-to sets have been
optimised to avoid unnecessary copying.
Field-sensitive constant propagation for structures and cell-sensitive
constant propagation for arrays have been introduced as well as
propagation of conditions has been introduced in order to filter
points-to sets and avoid exploration of unfeasible branches.

\begin{figure}[t]
  \includegraphics[width=\textwidth]{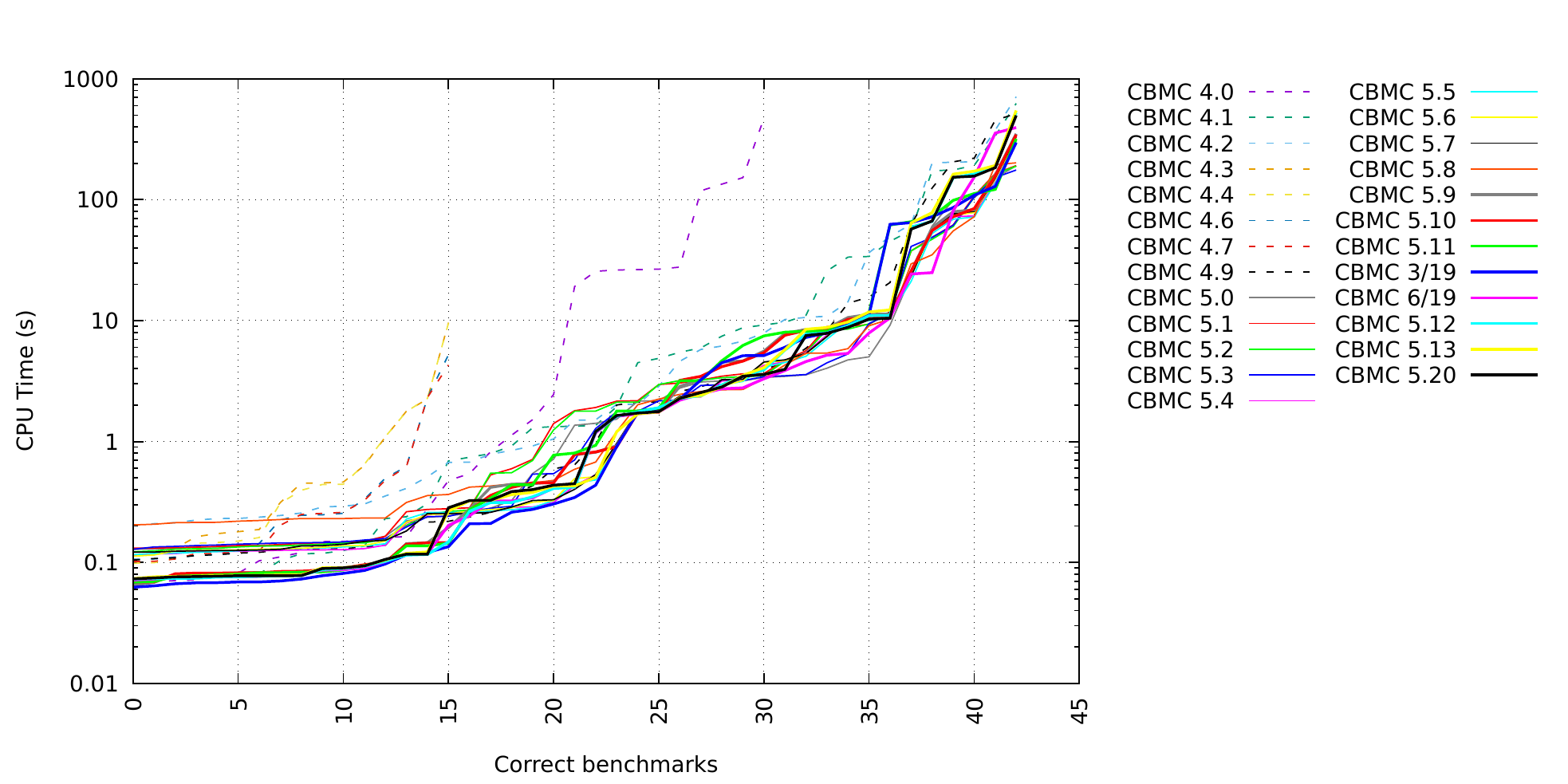}
  \caption{CPU time quantile plot over ReachSafety-Floats
  }
  \label{fig:cpu-cactus-ReachSafety-Floats}
\end{figure}

Figure~\ref{fig:cpu-cactus-ReachSafety-Floats} shows the evolution of CPU time
on the ReachSafety-Floats benchmarks.
CBMC's floating point decision procedure has seen a sustained period
of bug fixes and optimisations, in particular between CBMC 4.2 and
5.0.
%
%
CBMC 5.8 introduced a more complete built-in library for
\texttt{math.h}, which explains that later 5.x versions solve many more
benchmarks than the earlier ones.

Overall, we observe that the performance evolution of CBMC was
not so much dominated by a few major features that gave a massive
performance boost, but rather improved through a steady stream of
incremental enhancements and bug fixes.

\section{Future Directions}\label{sec:future}

CBMC has proven to be able to verify and find bugs
in real, large-scale software projects.
Despite successes~\cite{DBLP:journals/spe/ChongCEKKMSTTT21},
the use of a software verification tool in such a context
is far from an easy task that often requires expert users.
Moreover, it requires a significant
amount of manual work to divide and conquer the application
in a modular way and writing harnesses and stub functions
with realistic assumptions about the environment the
program is executing in.
Hence, besides the perpetual endeavours of improving CBMC's
performance, improving the usability of the tool for
CBMC users is the main focus of development.


\begin{thebibliography}{10}
\providecommand{\url}[1]{{#1}}
\providecommand{\urlprefix}{URL }
\expandafter\ifx\csname urlstyle\endcsname\relax
  \providecommand{\doi}[1]{DOI~\discretionary{}{}{}#1}\else
  \providecommand{\doi}{DOI~\discretionary{}{}{}\begingroup
  \urlstyle{rm}\Url}\fi

\bibitem{DBLP:conf/cav/AlglaveKT13}
Alglave, J., Kroening, D., Tautschnig, M.: Partial orders for efficient bounded
  model checking of concurrent software.
\newblock In: {CAV}, \emph{LNCS}, vol. 8044, pp. 141--157 (2013)

\bibitem{DBLP:conf/popl/AlpernWZ88}
Alpern, B., Wegman, M.N., Zadeck, F.K.: Detecting equality of variables in
  programs.
\newblock In: POPL, pp. 1--11 (1988)

\bibitem{ANSI:1999:AII}
{American National Standards Institute}: {ANSI\slash ISO\slash IEC 9899-1999}:
  Programming Languages --- {C}.
\newblock American National Standards Institute, 1430 Broadway, New York, NY
  10018, USA (1999)

\bibitem{BarrettSST09}
Barrett, C., Sebastiani, R., Seshia, S.A., Tinelli, C.: Satisfiability Modulo
  Theories, \emph{Frontiers in Artificial Intelligence and Applications}, vol.
  185, chap.~26, pp. 825--885.
\newblock IOS Press (2009)

\bibitem{DBLP:conf/cav/BarrettCDHJKRT11}
Barrett, C.W., Conway, C.L., Deters, M., Hadarean, L., Jovanovic, D., King, T.,
  Reynolds, A., Tinelli, C.: {CVC4}.
\newblock In: G.~Gopalakrishnan, S.~Qadeer (eds.) Computer Aided Verification -
  23rd International Conference, {CAV} 2011, Snowbird, UT, USA, July 14-20,
  2011. Proceedings, \emph{Lecture Notes in Computer Science}, vol. 6806, pp.
  171--177. Springer (2011).
\newblock \doi{10.1007/978-3-642-22110-1\_14}.
\newblock \urlprefix\url{https://doi.org/10.1007/978-3-642-22110-1\_14}

\bibitem{BCCZ99}
Biere, A., Cimatti, A., Clarke, E.M., Zhu, Y.: Symbolic model checking without
  {BDDs}.
\newblock In: Tools and Algorithms for the Construction and Analysis of
  Systems, \emph{Lecture Notes in Computer Science}, vol. 1579, pp. 193--207.
  Springer (1999)

\bibitem{handbook09}
Biere, A., Heule, M., van Maaren, H., Walsh, T. (eds.): Handbook of
  Satisfiability, \emph{Frontiers in Artificial Intelligence and Applications},
  vol. 185. {IOS} Press (2009)

\bibitem{DBLP:journals/spe/ChongCEKKMSTTT21}
Chong, N., Cook, B., Eidelman, J., Kallas, K., Khazem, K., Monteiro, F.R.,
  Schwartz{-}Narbonne, D., Tasiran, S., Tautschnig, M., Tuttle, M.R.:
  Code-level model checking in the software development workflow at amazon web
  services.
\newblock Softw. Pract. Exp. \textbf{51}(4), 772--797 (2021).
\newblock \doi{10.1002/spe.2949}.
\newblock \urlprefix\url{https://doi.org/10.1002/spe.2949}

\bibitem{DBLP:conf/tacas/ClarkeKL04}
Clarke, E.M., Kroening, D., Lerda, F.: A tool for checking {ANSI-C} programs.
\newblock In: {TACAS}, \emph{LNCS}, vol. 2988, pp. 168--176 (2004)

\bibitem{DBLP:conf/dac/ClarkeKY03}
Clarke, E.M., Kroening, D., Yorav, K.: Behavioral consistency of c and verilog
  programs using bounded model checking.
\newblock In: DAC, pp. 368--371. ACM (2003)

\bibitem{DBLP:conf/fmcad/CookDKMPPTW20}
Cook, B., D{\"{o}}bel, B., Kroening, D., Manthey, N., Pohlack, M., Polgreen,
  E., Tautschnig, M., Wieczorkiewicz, P.: Using model checking tools to triage
  the severity of security bugs in the xen hypervisor.
\newblock In: 2020 Formal Methods in Computer Aided Design, {FMCAD} 2020,
  Haifa, Israel, September 21-24, 2020, pp. 185--193. {IEEE} (2020).
\newblock \doi{10.34727/2020/isbn.978-3-85448-042-6\_26}.
\newblock
  \urlprefix\url{https://doi.org/10.34727/2020/isbn.978-3-85448-042-6\_26}

\bibitem{DBLP:conf/cav/CookKKTTT18}
Cook, B., Khazem, K., Kroening, D., Tasiran, S., Tautschnig, M., Tuttle, M.R.:
  Model checking boot code from {AWS} data centers.
\newblock In: Computer Aided Verification - 30th International Conference,
  {CAV} 2018, Held as Part of the Federated Logic Conference, FloC 2018,
  Oxford, UK, July 14-17, 2018, Proceedings, Part {II}, \emph{Lecture Notes in
  Computer Science}, vol. 10982, pp. 467--486. Springer (2018).
\newblock \doi{10.1007/978-3-319-96142-2\_28}.
\newblock \urlprefix\url{https://doi.org/10.1007/978-3-319-96142-2\_28}

\bibitem{CKKST18}
Cordeiro, L.C., Kesseli, P., Kroening, D., Schrammel, P., Trt{\'{\i}}k, M.:
  {JBMC:} {A} bounded model checking tool for verifying {Java} bytecode.
\newblock In: Computer Aided Verification, {CAV}, \emph{LNCS}, vol. 10981, pp.
  183--190. Springer (2018)

\bibitem{DBLP:conf/popl/CytronFRWZ89}
Cytron, R., Ferrante, J., Rosen, B.K., Wegman, M.N., Zadeck, F.K.: An efficient
  method of computing static single assignment form.
\newblock In: POPL, pp. 25--35 (1989)

\bibitem{DBLP:series/txtcs/KroeningS16}
Kroening, D., Strichman, O.: Decision Procedures - An Algorithmic Point of
  View, Second Edition.
\newblock Texts in Theoretical Computer Science. An {EATCS} Series. Springer
  (2016).
\newblock \doi{10.1007/978-3-662-50497-0}.
\newblock \urlprefix\url{https://doi.org/10.1007/978-3-662-50497-0}

\bibitem{DBLP:conf/memics/KroeningT14}
Kroening, D., Tautschnig, M.: Automating software analysis at large scale.
\newblock In: MEMICS, \emph{Lecture Notes in Computer Science}, vol. 8934, pp.
  30--39. Springer (2014).
\newblock \doi{10.1007/978-3-319-14896-0_3}.
\newblock \urlprefix\url{http://dx.doi.org/10.1007/978-3-319-14896-0_3}

\bibitem{DBLP:conf/isola/MalacariaTD16}
Malacaria, P., Tautschnig, M., Distefano, D.: Information leakage analysis of
  complex {C} code and its application to {OpenSSL}.
\newblock In: ISoLA, \emph{Lecture Notes in Computer Science}, vol. 9952, pp.
  909--925 (2016).
\newblock \doi{10.1007/978-3-319-47166-2_63}.
\newblock \urlprefix\url{http://dx.doi.org/10.1007/978-3-319-47166-2_63}

\bibitem{DBLP:conf/tacas/MouraB08}
de~Moura, L.M., Bj{\o}rner, N.: {Z3:} an efficient {SMT} solver.
\newblock In: C.R. Ramakrishnan, J.~Rehof (eds.) Tools and Algorithms for the
  Construction and Analysis of Systems, 14th International Conference, {TACAS}
  2008, Held as Part of the Joint European Conferences on Theory and Practice
  of Software, {ETAPS} 2008, Budapest, Hungary, March 29-April 6, 2008.
  Proceedings, \emph{Lecture Notes in Computer Science}, vol. 4963, pp.
  337--340. Springer (2008).
\newblock \doi{10.1007/978-3-540-78800-3\_24}.
\newblock \urlprefix\url{https://doi.org/10.1007/978-3-540-78800-3\_24}

\bibitem{DBLP:conf/wcet/NemerCSBM06}
Nemer, F., Cass{\'{e}}, H., Sainrat, P., Bahsoun, J.P., Michiel, M.D.:
  Papabench: a free real-time benchmark.
\newblock In: 6th Intl. Workshop on Worst-Case Execution Time {(WCET)}
  Analysis, July 4, 2006, Dresden, Germany, \emph{{OASICS}}, vol.~4 (2006).
\newblock \urlprefix\url{http://drops.dagstuhl.de/opus/volltexte/2006/678}

\bibitem{DBLP:conf/lpar/NieuwenhuisOT04}
Nieuwenhuis, R., Oliveras, A., Tinelli, C.: Abstract {DPLL} and abstract {DPLL}
  modulo theories.
\newblock In: LPAR, \emph{Lecture Notes in Computer Science}, vol. 3452, pp.
  36--50. Springer (2004)

\bibitem{DBLP:journals/jacm/NieuwenhuisOT06}
Nieuwenhuis, R., Oliveras, A., Tinelli, C.: Solving sat and sat modulo
  theories: From an abstract davis--putnam--logemann--loveland procedure to
  dpll({\it }).
\newblock J. ACM \textbf{53}(6), 937--977 (2006)

\bibitem{DBLP:conf/popl/RosenWZ88}
Rosen, B.K., Wegman, M.N., Zadeck, F.K.: Global value numbers and redundant
  computations.
\newblock In: POPL, pp. 12--27 (1988)

\bibitem{UNIX98:LP64}
{The Open Group}: Data Size Neutrality and 64-bit Support.
\newblock IEEE (1998)

\end{thebibliography}

\end{document}